\documentclass[preprint]{aastex}
\slugcomment{Accepted by ApJ}

\shorttitle{TYPE IA SN PHYSICS FROM THE TYCHO SNR}
\shortauthors{Badenes et al.}

\begin{document}

\title{Constraints on the Physics of Type Ia Supernovae from the X-Ray Spectrum of the Tycho Supernova Remnant}

\author{Carles Badenes\altaffilmark{1}, Kazimierz J. Borkowski\altaffilmark{2}, John P. Hughes\altaffilmark{1}, Una
  Hwang\altaffilmark{3} and Eduardo Bravo\altaffilmark{4}}

\altaffiltext{1}{Department of Physics and Astronomy, Rutgers University, 136 Frelinghuysen Rd., 
  Piscataway NJ 08854-8019; badenes@physics.rutgers.edu; jph@physics.rutgers.edu}

\altaffiltext{2}{Department of Physics, North Carolina State University, Box 8202, Raleigh NC 27965-8202;
  kborkow@unity.ncsu.edu}

\altaffiltext{3}{NASA Goddard Space Flight Center, Code 622, Greenbelt MD 20771; and Department of Physics and 
  Astronomy, The Johns Hopkins University, 3400 Charles St, Baltimore MD 21218; hwang@orfeo.gsfc.nasa.gov}

\altaffiltext{4}{Departament de F\'{i}sica i Enginyeria Nuclear, Universitat Polit\`{e}cnica de Catalunya, Diagonal 647,
  Barcelona 08028, Spain; and Institut d'Estudis Espacials de Catalunya, Campus UAB, Facultat de Ci\`{e}ncies. Torre
  C5. Bellaterra, Barcelona 08193, Spain; eduardo.bravo@upc.es}

\begin{abstract}
  In this paper we use high quality X-ray observations from \textit{XMM-Newton} and \textit{Chandra} to gain new
  insights into the explosion that originated Tycho's supernova 433 years ago. We perform a detailed comparison between
  the ejecta emission from the spatially integrated X-ray spectrum of the supernova remnant and current models for Type
  Ia supernova explosions. We use a grid of synthetic X-ray spectra based on hydrodynamic models of the evolution of the
  supernova remnant and nonequilibrium ionization calculations for the state of the shocked plasma. We find that the
  fundamental properties of the X-ray emission in Tycho are well reproduced by a one-dimensional delayed detonation
  model with a kinetic energy of $\sim1.2\cdot10^{51}$ erg. All the other paradigms for Type Ia explosions that we have
  tested fail to provide a good approximation to the observed ejecta emission, including one-dimensional deflagrations,
  pulsating delayed detonations and sub-Chandrasekhar explosions, as well as deflagration models calculated in three
  dimensions. Our results require that the supernova ejecta retain some degree of chemical stratification, with Fe-peak
  elements interior to intermediate mass elements. This strongly suggests that a supersonic burning front (i.e., a
  detonation) must be involved at some stage in the physics of Type Ia supernova explosions.
\end{abstract}


\section{INTRODUCTION} \label{sec:Intro}

The \textit{stella nova} of 1572 is the only historical supernova (SN) that can be classified with some degree of
confidence as Type Ia based on its observed light curve and color evolution \citep{ruiz-lapuente04:TychoSN}. Our
knowledge of the universe has improved greatly since this momentous event was recorded by Tycho Brahe and other
astronomers in the sixteenth century, but now, as then, there are many unanswered questions regarding the nature of
supernovae. Type Ia SNe play a key role as cosmological probes
\citep{riess98:accelerating_universe,perlmutter99:omega_lambda_supernovae}, and they constitute the most direct
evidence for the accelerating universe \citep{leibungut01:type_ia_review2}, but despite the continuing efforts of
theorists during the last decades, our understanding of these objects is far from being complete. There is a general
agreement that Type Ia SNe are the result of the thermonuclear explosion of a C+O white dwarf (WD) that is destabilized
by accretion from a companion star in a close binary system. This so-called `single degenerate' scenario has been
recently substantiated by the discovery of a G-type star near the center of the Tycho supernova remnant (SNR) that
appears to be the surviving companion to the WD that exploded as a SN in 1572 \citep{ruiz-lapuente04:Tycho_Binary}, and
by the presence of a weak H$\alpha$ signature in the late time spectrum of the unusual Type Ia SN 2002ic
\citep{hamuy03:agb}. However, the fundamental details of the explosion itself are still obscure, and a number of
contending models or paradigms are currently being considered.

Most of these paradigms assume that the WD becomes unstable and then explodes when its mass approaches the Chandrasekhar
limit ($1.4\,\mathrm{M_{\odot}}$). In this case, the properties of the explosion are determined mainly by the
propagation mode of the burning front through the star, which can be supersonic (detonations), subsonic (deflagrations),
or a combination of both (delayed detonations, pulsating delayed detonations). In the prompt detonation models, the
burning front propagates supersonically through the star, incinerating almost its entire mass to Fe-peak nuclei. In pure
deflagration models, the burning front propagates subsonically and the star has the time to expand ahead of the flame.
When the expansion velocity of the unburnt material becomes comparable to that of the burning front, the flame quenches,
leaving a mixture of unburnt fuel (C and O) and ashes (Fe-peak nuclei) in the ejecta, usually with trace amounts of
intermediate mass elements (IMEs: Si, S, Ar, Ca, etc.). In one-dimensional deflagration models, the ejecta are layered,
with the ashes interior to the fuel, but three-dimensional calculations have shown that the turbulent properties of the
burning front lead to an efficient mixture of fuel and ashes through the ejecta. In the delayed detonation models, the
burning front begins as a deflagration and then it is artificially forced to make the transition to a detonation,
usually when a prescribed transition density $\rho_{tr}$ is reached. In this kind of explosions, most of the WD is burnt
and very little C and O is left behind, but the expansion of the star during the initial deflagration allows for the
detonation to burn a sizable mass of ejecta at intermediate densities, leading to a significant production of IMEs.
Pulsating delayed detonations are a variation of this scenario, where the flame quenches after a brief deflagration
phase and the WD re-collapses, triggering a detonation in the process. The ejecta are stratified in all the flavors of
delayed detonations simulated in one dimension, with Fe-peak nuclei interior to IMEs, and some unburnt C and O in the
outermost layers.  Whether this is also the case for three-dimensional delayed detonations is still under debate. Given
the appropriate circumstances, it is also possible for the WD to explode with a sub-Chandrasekhar mass, leading to an
entirely different class of models whose initial conditions are not so well determined. Sub-Chandrasekhar explosions
usually involve the ignition under degenerate conditions of a layer of accreted He on top of the C+O WD, which sends a
strong pressure wave towards the center of the star that eventually triggers a detonation. The structure of the ejecta
is complex, with unburnt C and O and IMEs sandwiched between two regions rich in Fe-peak nuclei. For more details on
these explosion paradigms and a complete set of references, see the reviews by \citet{hillebrandt00:Ia-review} (mostly
one-dimensional models), \citet{bravo04:3D_review} (three-dimensional deflagrations), and the recent works by
\citet{garcia-senz03:DDT3D} and \citet{gamezo05:DDT3D} (three-dimensional delayed detonations).

There have been numerous attempts to constrain these theoretical models, mostly through the optical/IR spectra
\citep[e.g.,][]{hoeflich96:nosubCh,fisher97:SN1990N,hoeflich98:TypeIa-initial-composition,wheeler98:TypeIa-diagnostics-IR,branch05:SN1994D,kozma05:3dIa_late_time_spectra}
and light curves \citep[e.g.,][]{hoeflich98:TypeIa-initial-composition,stritzinger05:Typeia_constraints} of Type Ia SNe,
and also through the few available gamma-ray observations \citep[][and references therein]{milne04:Typeia_gamma_ray}.
Although some preference has been shown for delayed detonation models in these studies, it cannot be said that a general
consensus has been reached in this matter, and the physical mechanism responsible for Type Ia SN explosions still
remains an open issue.  In the present paper, we address the problem of constraining the physics of Type Ia SN
explosions from a different perspective.  Instead of focusing on the emission from the SNe themselves, we take advantage
of the excellent quality of the existing X-ray observations of SNRs provided by \textit{XMM-Newton} and \textit{Chandra}
to probe the physical mechanism responsible for the explosions. The necessary theoretical groundwork was laid down in
\citet{badenes03:xray} and \citet{badenes05:xray} (henceforth, Paper I and Paper II), where we showed that the structure
of the SN ejecta has a profound impact on the thermal X-ray emission from Type Ia SNRs. As a consequence, the X-ray
spectra have the potential to pose strong constraints on the kind of explosion that originated the SNR. Here we apply
the models developed in Papers I and II to the X-ray spectrum of the Tycho SNR. We begin by summarizing the observations
of this object in \S~\ref{sec:Observations}. Based on the observational properties of Tycho, we outline our strategy for
comparing our models to the observations in \S~\ref{sec:Goals-and-strategy}. In \S~\ref{sec:Line-emission}, we extract
spatially integrated spectra from large regions of an \textit{XMM-Newton} observation of Tycho and derive the
fundamental properties of the line emission from the shocked SN ejecta. We compare our models with these observations in
two stages: first we select the most promising models based on the properties of their line emission in
\S~\ref{sec:Line-emission-models} and then we compare the best candidates to the spatially integrated spectrum in
\S~\ref{sec:Modeling-Integrated_Spectrum}.  In \S~\ref{sec:Spatial} we consider the spatial distribution of the line
emission from a \textit{Chandra} ACIS-I observation, and examine the implications for our models.  Finally, we discuss
the performance of the models from a global point of view in \S~\ref{sec:discussion}, and we summarize our conclusions
in \S~\ref{sec:Conclusions}.

\section{OBSERVATIONS} \label{sec:Observations}

\subsection{The supernova of 1572} \label{sub:Obs-Tycho-SN}

As early as 1945, \citeauthor{baade45:SN1572} used the observations made and compiled by Tycho Brahe to derive a light
curve that enabled him to identify the \textit{stella nova} of 1572 as a Type I supernova \citep{baade45:SN1572}.  Since
then, the classification of the supernova as determined from its light curve has been controversial, and some authors
have claimed that SN 1572 was a subluminous event \citep[see][and references
therein]{vandenBergh93:Tycho_SN_subluminous?,schaefer96:Historical_SNe}.  \citet{ruiz-lapuente04:TychoSN} performed a
detailed reanalysis of the sixteenth century records and evaluated the uncertainties in the data in order to describe
the light curve in the terms used nowadays to characterize Type Ia SNe. According to
\citeauthor{ruiz-lapuente04:TychoSN}, SN 1572 was a normal Type Ia SN within the uncertainties associated with the data
(which are large), with a peak visual magnitude of $\mathcal{M}_{V}=-19.24-5log(D/3.0 \mathrm{kpc})\pm0.42$ mag, and a
stretch factor \footnote{The stretch factor $s$ is a parameterization of the light curve width/shape relationship,
  defined as the linear broadening or narrowing of the rest-frame time scale of an average template light curve that is
  required to match the observed light curve. Values of $s$ close to 1 denote `normal' Type Ia SNe. See
  \citet{perlmutter97:omega_lambda} for more details.}  $s \backsim 0.9$. The author found an extinction of
$A_{V}=1.86\pm0.12$ mag and an average reddening of $E(B-V)=0.6\pm0.04$ in that direction of the sky, which combined
with the value of $\mathcal{M}_{V}$ yield an estimate of $2.8\pm0.4$ kpc for the distance to the SN, assuming
$H_{0}=65\, \mathrm{km \cdot s^{-1} \cdot Mpc^{-1}}$.  

\subsection{The Tycho SNR: Radio and optical observations}\label{sub:Obs-radio-optical}

In the radio continuum observations at 1375 MHz ($\lambda=21$ cm), the Tycho SNR appears as a clearly defined shell with
an approximate angular diameter of 8'. The shell is nearly spherical from the northwest to the southeast, with an
irregular outbreak accompanied by a slight brightening to the north, northeast and east. \citet{reynoso97:Tycho_VLA},
used VLA observations to estimate the expansion parameter of the forward shock, $\eta_{fwd}$ (defined as
$\eta=d\mathrm{log}(r_{fwd})/d\mathrm{log}(t)$), along the rim of the SNR shell.  They found an average of
$\eta_{fwd}=0.47\pm0.03$, with distinctly lower values towards the north and east. It was found later that the lower
value of $\eta_{fwd}$ in the northeast is due to an interaction with a dense cloud of neutral hydrogen
\citep{reynoso99:Tycho_VLA_II,lee04:envir_Tycho}. There is a puzzling, and still unsolved, disagreement between these
radio measurements and the value of $\eta_{fwd}=0.71\pm0.06$ inferred from X-ray observations by
\citet{hughes00:Expansion_Tycho_X_Rays}.  The expansion parameters of several interior features, measured using the same
techniques, are consistent in radio \citep[$\eta\simeq0.44$,][]{reynoso97:Tycho_VLA} and X-rays
\citep[$\eta\simeq0.45$,][]{hughes00:Expansion_Tycho_X_Rays}.

At optical wavelengths, only a few faint filaments of Balmer line emission from H are visible at the outer rim of the
SNR. No evidence for any optical emission other than the Balmer H lines was found in the interior of the SNR
\citep{kirshner78:tycho_SNR,smith91:six_balmer_snrs,ghavamian00:shock_precursor_Tycho}. In particular, optical emission
from radiatively cooled ejecta has not been detected. In \citet{ghavamian01:balmer_SNRs}, the spectrum of
the brightest knot in the eastern rim \citep[knot $g$ from][]{kamper78:optical_expansion_Tycho} was examined in
detail, yielding a velocity between $1940$ and $2300\, \mathrm{km\cdot s^{-1}}$ for the forward shock. This bright knot,
however, is situated in the eastern rim, where the SNR is interacting with denser material, and its properties might not
be representative of the overall dynamics of the blast wave.

As is often the case with Galactic SNRs, the distance to Tycho is very uncertain. Different techniques yield different
and even contradictory results ranging anywhere between 1.5 and 4.5 kpc, with most estimates converging around 2.5 kpc
\citep[see][for a review]{schaefer96:Historical_SNe}. Among these, we believe that the most reliable results are obtained
through the combination of velocity estimates from knots in the Balmer-dominated forward shock with optical proper
motions measured over long temporal baselines. We will adopt the range of 1.5-3.1 kpc obtained by
\citet{smith91:six_balmer_snrs} with this method as a conservative estimate for the distance, and we note that the value
of 2.8 kpc found by \citet{ruiz-lapuente04:TychoSN} for SN 1572 falls within this range.


\subsection{The Tycho SNR: X-ray observations}\label{sub:Obs-X-ray}

The most fundamental properties of the X-ray emission from the Tycho SNR have been known since the time of the
\textit{ASCA} satellite \citep{hwang97:tycho-ASCA,hwang98:tycho-Feemission}, but the quality of the available data has
increased dramatically thanks to \textit{XMM-Newton} \citep{decourchelle01:tycho-xmm} and \textit{Chandra}
\citep{hwang02:tycho-rim,warren05:Tycho}. Morphologically, the X-ray emission from the Tycho SNR has a shell-like
structure with a thin rim of high energy continuum emission. This rim is coincident in position, but not in brightness,
with the outer edge of the radio emission, and traces the position of the forward shock.  The regions interior to this
rim show very strong emission lines of Si, S, Ca, Ar and Fe, which are thought to arise from the shocked supernova
ejecta.

Several significant results about the dynamics and X-ray emission of the Tycho SNR were presented in
\citet{warren05:Tycho}, where a principal component analysis technique was applied to a $\sim 150$ ks \textit{Chandra}
ACIS-I observation. \citeauthor{warren05:Tycho} determined the positions of the forward shock (FS), reverse shock (RS)
and contact discontinuity (CD) between ejecta and AM, and derived ratios for their average radii of 1:0.93:0.71
(FS:CD:RS). Because the CD is so close to the FS, these ratios are incompatible with one-dimensional adiabatic
hydrodynamics with $\gamma=5/3$ for the shocked AM. Three explanations for the presence of ejecta close to the FS can be
found in the literature: the presence of clumps with a high density contrast in the ejecta \citep{wang01:Ia-inst-clump},
the effect of cosmic ray (CR) acceleration at the FS \citep{decourchelle00:cr-thermalxray,blondin01:RT_instab_SNR_CR},
and the interaction of Rayleigh-Taylor fingers with circumstellar cloudlets \citep{jun96:rt_fingers}. The last mechanism
is unlikely to be relevant in the case of the Tycho SNR, because there is no known process that would produce an almost
isotropic distribution of cloudlets in the circumstellar environment of a Type Ia SN progenitor \citep[see Figure 4
in][for the azimuthal distribution of FS and CD radii]{warren05:Tycho}. According to \citeauthor{warren05:Tycho}, the
absence of prominent variations in the spectral properties of the ejecta along the CD seems to favor CR acceleration
over ejecta clumping as the dominating mechanism.

\citeauthor{warren05:Tycho} also found that the properties of the X-ray emission from the shocked AM are compatible with
a CR-dominated blast wave. On one hand, the emitting region behind the FS is much thinner than would be predicted by the
thermal emission behind an adiabatic shock \citep[see also][]{ballet05:x-ray-synchrotron}. On the other hand, the
featureless spectra extracted from the shocked ambient medium (AM) were well fitted by a power law with an index of
$\sim 2.7$. This value is consistent with the index of $2.72$ found in the high energy \textit{Ginga} observations by
\citet{fink94:tycho-ginga}.  Fits with thermal models at $kT\simeq2\,\mathrm{keV}$ are also possible, but they require
extremely low values for the ionization timescale \citep[a few times $10^{8}\,\mathrm{cm^{-3}\cdot
  s}$,][]{hwang02:tycho-rim}, which are incompatible with the basic properties of the Tycho SNR \citep[see the
discussion in \S~7.4.1 of][]{warren05:Tycho}

The emission from the shocked ejecta, as seen in the spectral bands corresponding to the most prominent emission lines
from heavy elements, has a shell-like morphology, with the Fe K$\alpha$ line image appearing more diffuse and peaking at
a smaller radius than the others \citep{hwang97:tycho-ASCA, decourchelle01:tycho-xmm}. The Fe K$\alpha$ emission is not
only spatially distinct from the other lines (including the Fe L complex); it also has different spectral properties,
with a higher temperature and a lower ionization timescale \citep{hwang98:tycho-Feemission}. The apparent symmetry of
the X-ray line emission and the absence of significant Doppler shifts suggests an overall spherical geometry, but local
inhomogeneities in the line emission are manifest, particularly in the form of bright clumps in the southeast
\citep[see][]{vancura95:Tycho, decourchelle01:tycho-xmm}.

\section{GOALS AND STRATEGY}\label{sec:Goals-and-strategy}

The goal of this paper is to model the thermal X-ray emission from the shocked ejecta in the Tycho SNR using the grid of
synthetic spectra presented in Papers I and II. Our objective is to use these synthetic spectra to constrain the physics
of the Type Ia SN explosion that gave birth to the remnant in 1572. We shall make no attempt to model the featureless
emission from the shocked AM.  Instead, we adopt the view expressed in \citet{warren05:Tycho} that this emission is
predominantly nonthermal, and that the dynamics of the FS are strongly modified by CR acceleration. Although the
one-dimensional hydrodynamic calculations with $\gamma=5/3$ that underlie our synthetic spectra cannot reproduce the
properties of the shocked AM or the dynamics of the FS under these conditions, this is not a serious drawback. Several
fundamental quantities like the temperature of the shocked AM or the expansion parameter of the FS are not known with a
significant degree of accuracy (see the discussions in \S~\ref{sub:Obs-radio-optical} and \S~\ref{sub:Obs-X-ray}).
Rather than use these poorly determined quantities to constrain the hydrodynamic models underlying our synthetic
spectra, we attempt to reproduce the observed X-ray emission first and then verify that the SNR dynamics is compatible
with the known properties of the AM and FS.

Regardless of the conditions at the FS, we believe that our hydrodynamic models with $\gamma=5/3$ are a good
approximation for the dynamics of the shocked ejecta. Diffusive shock acceleration is mediated by strong magnetic
fields, and leads to high compression ratios and low temperatures in the shocked plasma
\citep[see][]{ellison04:hd+cr,ellison05:hd+cr_revshock}. From an observational point of view, both the position of the
RS measured by \citet{warren05:Tycho}, deep into the ejecta and away from the CD, and the strong Fe K$\alpha$ emission
that originates from the hot plasma just behind the RS are strong arguments against efficient CR acceleration at the RS.
From a theoretical point of view, it is hard to see how the necessary magnetic fields could have survived for hundreds
of years inside the freely expanding SN ejecta, even making the most optimistic assumptions about the initial magnetic
field strength in the progenitor systems of Type Ia SNe.

As we discussed in Paper II (\S~A.2), a detailed comparison between our synthetic spectra and observations is not
straightforward. On one hand, the uncertainties in the atomic data and other factors do not make the model spectra
amenable to the usual $\chi^{2}$ fitting techniques, and on the other hand, the parameter space is very large even for
one-dimensional calculations. In the present work, we solve these problems by taking advantage of the characteristics of
the X-ray emission from the Tycho SNR to devise an efficient comparison strategy. First, we extract spatially integrated
spectra from large regions in the SNR, and determine the fundamental properties of the high-energy ($>1.5$ keV) line
emission (\S~\ref{sec:Line-emission}). As we have seen in \S~\ref{sub:Obs-X-ray}, this line emission comes entirely from
the shocked ejecta, so we use the observed line flux ratios and line centroids to reduce the dimensionality of the
problem. We require that our synthetic X-ray spectra reproduce as many of these flux ratios and centroids as possible,
and in the process we identify the most successful Type Ia SN explosion models and the most promising regions of the
associated parameter space (\S~\ref{sec:Line-emission-models}). Finally, we compare the best models to the entire
spatially integrated spectrum, applying several consistency checks to the requirements of each model, such as
interstellar absorption, AM flux, and derived distance estimates (\S~\ref{sec:Modeling-Integrated_Spectrum}).

\section{SPATIALLY INTEGRATED X-RAY SPECTRUM AND LINE EMISSION} \label{sec:Line-emission}

\begin{figure}

  \epsscale{0.6}

  \plotone{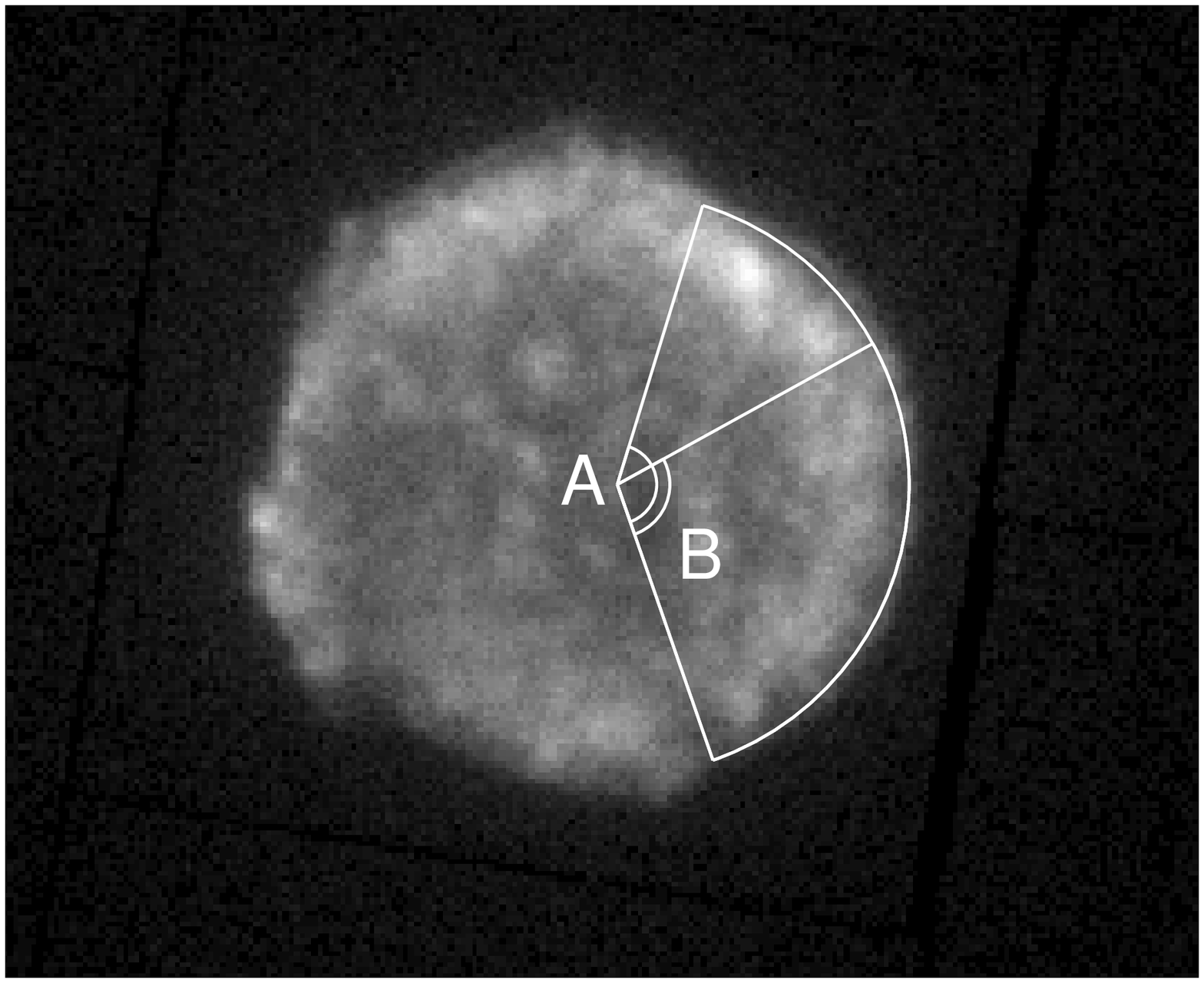}

  \caption{The \textit{XMM-Newton} EPIC MOS image of the Tycho SNR, with the spatial extraction regions A and B indicated. \label{fig-1}}

\end{figure}

\begin{figure}
  
  \epsscale{0.75}

  \plotone{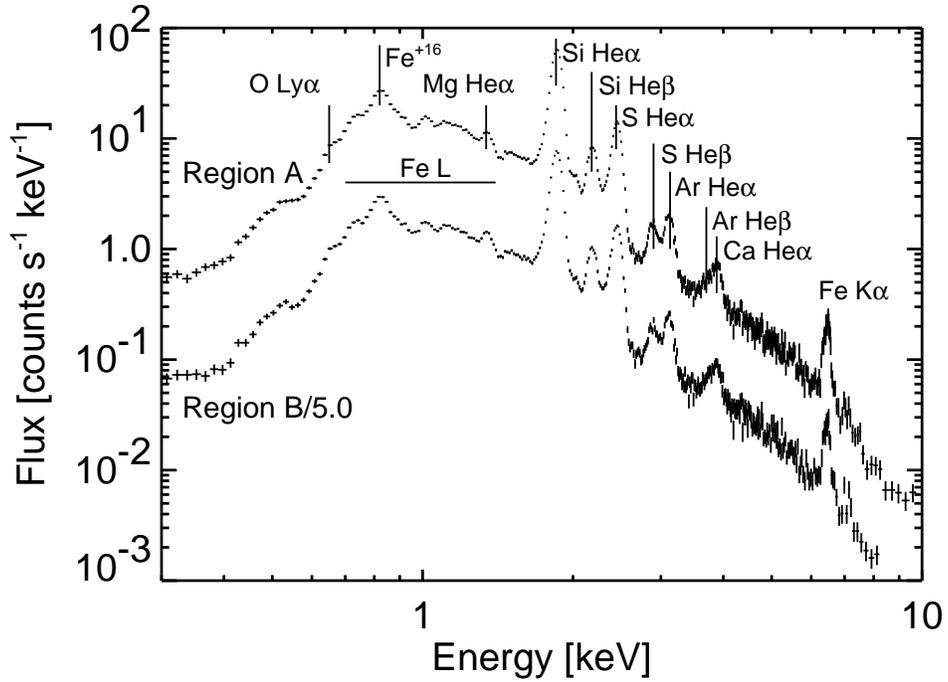}

  \caption{ \textit{XMM-Newton} EPIC MOS1 spectra of regions A and B. The spectrum of region B has been
    offset for clarity, dividing the flux by 5.0. The most important lines and line complexes have been marked.\label{fig-2}}

\end{figure}

Among the publicly available X-ray observations of the Tycho SNR, the best spatially integrated spectra are provided by
the EPIC MOS cameras on board \textit{XMM-Newton}. The observation analyzed in \citet{decourchelle01:tycho-xmm} has a
total exposure time of about 12 ks for each of the EPIC MOS CCDs, providing spectra of good quality for this bright
object \protect\footnote{The data are available from the \textit{XMM-Newton} science archive (XSA):
  http://xmm.vilspa.esa.es/external/xmm\_data\_acc/xsa/.}. Since our models are one-dimensional, we focus on the western
sector of Tycho, avoiding the prominent inhomogeneities in the ejecta emission found in the southeast and the
interaction with a denser AM towards the northeast (see discussion in \S~\ref{sub:Obs-radio-optical} and
\S~\ref{sub:Obs-X-ray}). In the west, the ejecta emission is homogeneous, and the FS has a smooth, almost circular
shape. Using the standard \textit{XMM-Newton} science analysis system, we have selected two spectral extraction regions
in the EPIC MOS1 image (see Fig. \ref{fig-1}). Region A covers the entire western sector between two major outbreaks in
the FS in the north and south; it corresponds loosely to region V in \citet{reynoso97:Tycho_VLA}, with a range in
position angle of $200 \leq \theta \leq 345$ degrees (counterclockwise from the north). Region B is a subset of region
A, covering the range $200 \leq \theta \leq 300$, and has been selected to exclude an area in the northwest that shows
an increased flux in the Fe and Si emission \citep[see Figs. 2 and 3 in][]{decourchelle01:tycho-xmm}. The spectra
extracted from regions A and B are shown in Figure \ref{fig-2}.

Many of the lines and line complexes in the spectra of Figure \ref{fig-2} are blended due to the limited spectral
resolution of CCD cameras in the X-ray band. In order to characterize the properties of the line emission in regions A
and B, we have fitted the extracted spectra between 1.65 and 9.0 keV with a model adapted from
\citet{hwang97:tycho-ASCA}, consisting of fourteen Gaussian lines plus two components for the continuum (a thermal
bremsstrahlung and a power law), attenuated by the interstellar absorption. For the neutral hydrogen column density, we
have adopted a value of $N_{H}=0.6 \cdot 10^{22}\,\mathrm{cm^{-2}}$, which is compatible with the
$0.55-0.59\cdot10^{22}\,\mathrm{cm^{-2}}$ range found by \citet{hwang98:tycho-Feemission} from fits to the integrated
ASCA spectrum. It is also close to the values determined by fitting the \textit{Chandra} spectra in several locations
along the western rim with nonthermal models: $0.58-0.69 \cdot 10^{22}\,\mathrm{cm^{-2}}$ \citep{warren05:Tycho};
$0.71-0.79\cdot10^{22}\,\mathrm{cm^{-2}}$ \citep{hwang02:tycho-rim}.  Conversion of the value of $A_{V}$ obtained by
\citet{ruiz-lapuente04:TychoSN} to $N_{H}$ with the $A_{V}$--$N_{H}$ relation of \citet{predehl95:Dust_Scattering}
yields a lower column density of $0.34 \cdot 10^{22}\,\mathrm{cm^{-2}}$, but the uncertainties associated with this
estimate are large.  We stress that the interstellar absorption in the X-ray band is hard to constrain in objects with
complex spectra such as SNRs, and variation of $N_{H}$ across the surface of an object as extended as Tycho cannot be
discarded (see \S~\ref{sub:comparing} for a discussion). For the power law index, we have adopted the value of 2.72
found by \citet{fink94:tycho-ginga} and confirmed by \citet{warren05:Tycho}.

\begin{figure}

  \centering

  \includegraphics[scale=0.6]{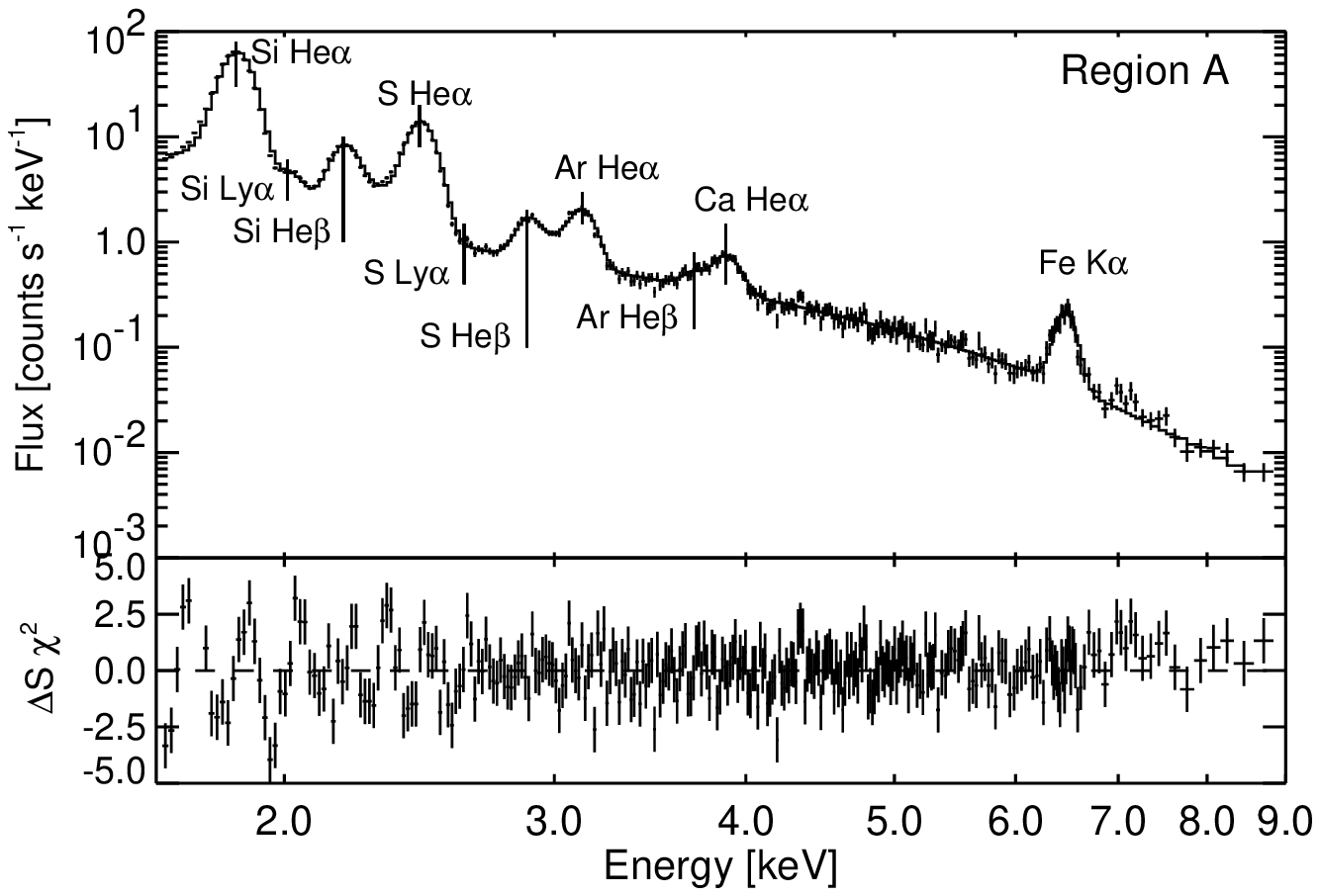}
  \includegraphics[scale=0.6]{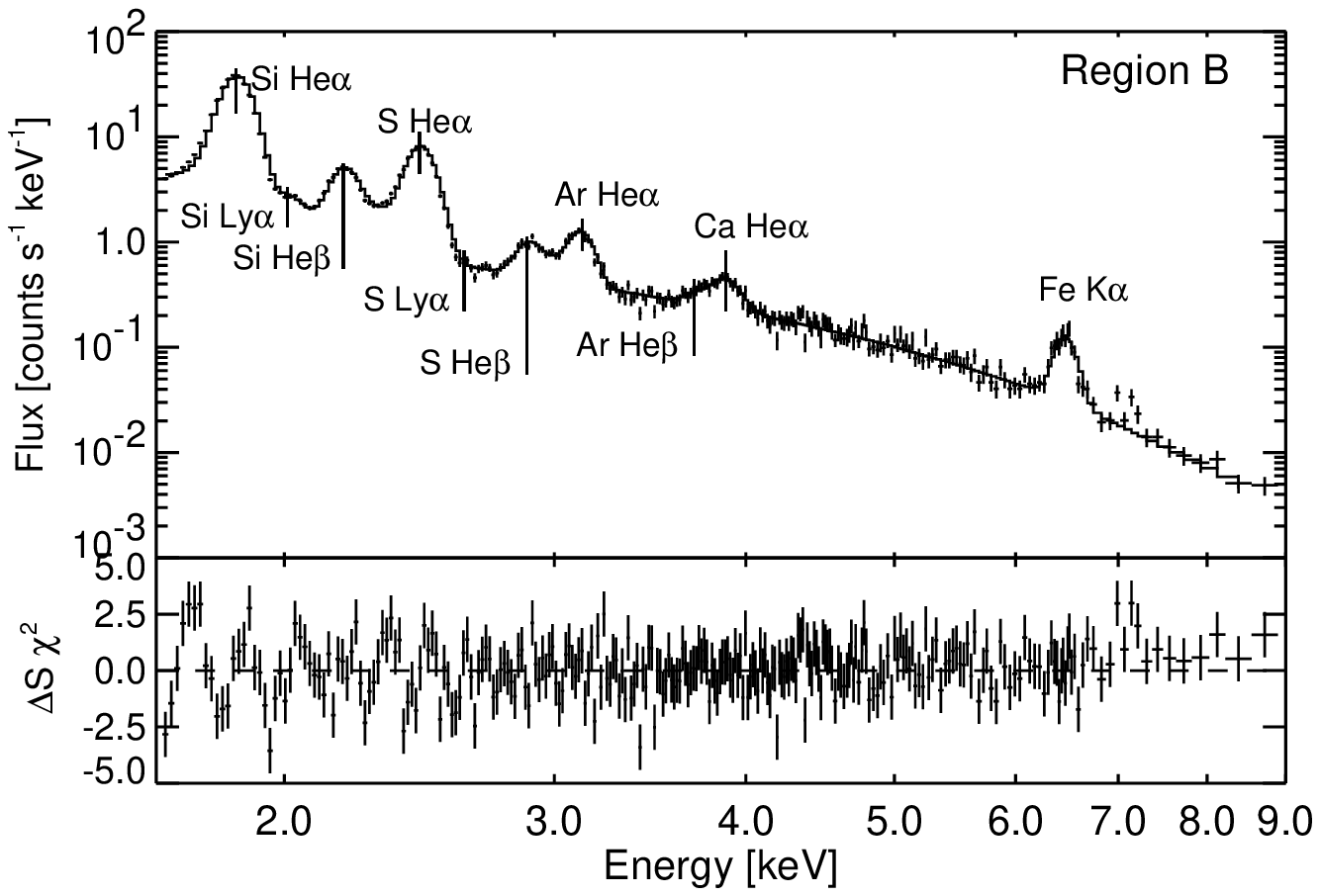}

  \caption{Fits to the line emission between 1.65 and 9 keV in the \textit{XMM-Newton} EPIC MOS1 spectrum of regions A and B. 
    The most important lines and line blends have been labeled for clarity.\label{fig-3}}

\end{figure}

The fits to regions A and B are shown in Figure \ref{fig-3}. The statistics for the fit to region A are
$\chi^{2}=462.04$ with $\chi^{2}/\mathrm{DOF}=1.52$, and for region B $\chi^{2}=373.75$ with
$\chi^{2}/\mathrm{DOF}=1.38$.  The fitted temperatures for the thermal bremsstrahlung component are 0.48 and 0.40 keV,
respectively. We note that these values are rather different from those obtained by \citet{hwang97:tycho-ASCA} assuming
a purely thermal continuum (these authors used a main component at 0.99 keV, with another at 10.0 keV to reproduce the
high energy continuum). The fourteen lines included in the model, along with the fitted centroids and fluxes, are given
in Table \ref{tab-1}, where the common notation of $\alpha$, $\beta$, and $\gamma$ has been used to label the lines
corresponding to transitions from quantum levels 2, 3, and 4 to level 1. The quality of the data set did not allow the
centroids of the weakest lines to be fitted independently, so these parameters were fixed.  The He$\gamma$/He$\beta$
line flux ratios of Si, S and Ar, and the Si Ly$\beta$/Si Ly$\alpha$ ratio, were also fixed in the fits to the values
listed in Table \ref{tab-1}, which correspond to the values at $T=10^{7}\, \mathrm{K}$ \footnote{Some of these flux
  ratios have been updated with respect to the values listed in \citet{hwang97:tycho-ASCA} using the atomic data from
  the ATOMDB compilation available at http://cxc.harvard.edu/atomdb/.}.  This allows for an adequate (if simple)
treatment of these blended lines, and is justified because the flux ratios vary by only 10\%-20\% over a decade in
temperature \citep[for details, see][]{hwang97:tycho-ASCA}. The most important line flux ratios are listed in Table
\ref{tab-2}. For these flux ratios, we have used the S He$\alpha$ blend as a reference instead of Si He$\alpha$ in order
to make their values less dependent on $N_{H}$.

An inspection of Tables \ref{tab-1} and \ref{tab-2} reveals that the properties of the line emission in the western half
of Tycho are remarkably uniform. Indeed, the only noticeable differences between the two extraction regions are in the
centroids of the Ar He$\alpha$, Ca He$\alpha$ and Fe K$\alpha$ blends, which are a few eV lower in region B (the
deviation is below 0.5\% in all cases, and within the 90\% confidence ranges for Ca He$\alpha$ and Fe K$\alpha$). This
suggests a lower ionization state for these elements in region B, but the origin of this lower ionization and its
relationship to the brightening in the ejecta emission in the northern part of region A is unclear. In the absence of a
plausible physical interpretation for these effects, we use region B as representative of the shocked ejecta emission in
the reminder of the paper, but we stress that the differences between regions A and B are very small. Our
\textit{XMM-Newton} results for the western half of Tycho are very similar to those obtained by
\citet{hwang97:tycho-ASCA} for the \textit{ASCA} observation of the entire SNR. Although the measurements were made in
different regions of the SNR, assuming different values for $N_{H}$, with different models for the continuum, and with
different instruments, the error bars overlap in almost all the line flux ratios and centroids, and the deviations are
never larger than a few percent. The only case where this is not true is the centroid of the Ca He$\alpha$ blend
\citep[3.879 and 3.867 keV in regions A and B vs. 3.818 keV in][]{hwang97:tycho-ASCA}. The neighboring Ar He$\beta$ line
is stronger in our fits, shifting the Ca He$\alpha$ centroid towards higher energies that are closer to the expected
value of 3.88 keV.
 
\section{MODELING THE LINE EMISSION} \label{sec:Line-emission-models}

\subsection{Synthetic spectra} \label{sub:preliminary}

The method to generate the synthetic X-ray spectra that we use to model the line emission from the Tycho SNR is
described in Papers I and II. For each Type Ia SN explosion model, we simulate the interaction of the ejecta with a
uniform AM using a one-dimensional hydrodynamic code. We assume that the plasma is a nonrelativistic monoatomic ideal
gas with $\gamma=5/3$. Then we calculate the nonequilibrium ionization (NEI) and associated processes for each fluid
element in the shocked ejecta, taking as inputs the temporal evolution of density and specific internal energy from the
hydrodynamic simulation and the chemical composition from the SN explosion model. To perform these calculations, we use
a method adapted from \citet{hamilton84:ejecta}, which takes into account important details that are relevant to the
specific case of a plasma dominated by heavy elements. In particular, the influence of the large number of electrons
ejected during the ongoing ionization on variables like the plasma temperature \citep[noted
by][]{brinkmann89:NEI_SNRs_Tycho} is dealt with in a self-consistent way.  Once the electron temperatures and ionization
states in all the shocked ejecta are calculated in this way, a spatially integrated X-ray spectrum is produced using an
appropriate spectral code.

This simulation scheme does not take into account radiative or ionization losses or thermal conduction in the shocked
plasma. \citet{hamilton84:ejecta} showed that ionization and radiative losses are of the same order in a plasma
dominated by heavy elements. In order to assess the importance of these processes, we applied the \textit{post-facto}
estimation of radiative losses described in section 3.5 of Paper I and section A.3 of Paper II to all the models
presented here. We found that radiative losses are unimportant for an object of the age of the Tycho SNR, with only a
few inconsequential exceptions that we will specify in \S~\ref{sub:Successful-models}. This is in agreement with both
the conspicuous absence of optically emitting ejecta (see \S~\ref{sub:Obs-radio-optical}) and the results of simulations
that do include radiative and ionization losses, like \citet{sorokina04:typeIasnrs}. These authors also showed that
efficient thermal conduction destroys any temperature gradients on timescales that are shorter than the age of the Tycho
SNR. Thus, the presence of efficient thermal conduction is incompatible with the spatial morphology of the Fe K$\alpha$
and Fe L emission in the Tycho SNR (see \S~\ref{sub:Obs-X-ray}), which implies that a temperature gradient must exist in
the shocked ejecta.

The synthetic spectra for the shocked SN ejecta are characterized by four parameters only: the Type Ia SN explosion
model used, the age of the SNR $t$, the density of the AM $\rho_{AM}$, and the ratio between the specific internal
energies of the electrons and ions at the reverse shock, $\beta = \varepsilon_{e}/\varepsilon_{i}$, which represents the
efficiency of collisionless electron heating. The dependence of the ejecta X-ray emission on each of these parameters is
discussed in Papers I and II.

The 32 Type Ia SN explosion models presented in Papers I and II include examples of all the paradigms described in
\S~\ref{sec:Intro}. Among these, we have selected a more convenient subgrid of 12 models: three delayed detonations
(DDTa, DDTc, and DDTe), three pulsating delayed detonations (PDDa, PDDc, and PDDe), three deflagrations (DEFa, DEFc, and
DEFf), one prompt detonation (DET), one sub-Chandrasekhar explosion (SCH), and one deflagration calculated in three
dimensions with a smooth particle hydrodynamics code \citep[B30U from][]{garcia-senz05:Ia3D}.  In order to
benchmark the results obtained using this subgrid of models, we have chosen to add three well-known Type Ia SN explosion
models calculated by other groups: model W7 \citep{nomoto84:W7}, a `fast deflagration' that has become a widespread
standard for comparison with observations of Type Ia SNe, model 5p0z22.25 \citep{hoeflich02:SN1999by}, a delayed
detonation calculated with a resolution 4 times greater than the DDT models of our grid, and model b30\_3d\_768
\citep{travaglio04:3D}, a deflagration calculated in three dimensions with an Eulerian code and a high spatial
resolution.

Of the three remaining parameters in the synthetic spectra, the age $t$ is known to be 433 years for the Tycho SNR (428
years at the time of the \textit{XMM-Newton} observation). Thus, only $\rho_{AM}$ and $\beta$ can be varied for each of
our Type Ia SN explosion models. We have sampled this parameter space with five points in $\rho_{AM}$ ( $2\cdot10^{-25}$
, $5\cdot10^{-25}$, $10^{-24}$, $2\cdot10^{-24}$, and $5\cdot10^{-24}\, \mathrm{g\cdot cm^{-3}}$ ) and three points in
$\beta$ ($\beta_{min}$\footnote{Defined as $\beta_{min}=\bar{Z_{s}}m_{e}/\bar{m_{i}}$, with $\bar{Z_{s}}$ the mean preshock
  ionization state, $m_{e}$ the electron mass and $\bar{m_{i}}$ the average ion mass, see \S~2.2 in paper II.}, $0.01$,
and $0.1$), setting $t$ to 430 yr in all the models. These ranges are selected to encompass the likely values of $\beta$
(see discussion in \S~2.2 of Paper II) and $\rho_{AM}$ in the Tycho SNR. For each Type Ia SN explosion model, 15
synthetic SNR spectra have been generated, for a total of 225 synthetic spectra.

\subsection{Comparing the models to the observed line emission} \label{sub:comparing}

We have selected thirteen observable parameters as diagnostic quantities for the comparison between our models and the
observations: the eight line flux ratios listed in Table \ref{tab-3} (He$(\beta+\gamma)$/He$\alpha$ and
Ly$\alpha$/He$\alpha$ ratios for Si and S, and the ratios of the He$\alpha$ blends of Si, Ar, and Ca, and the K$\alpha$
blend of Fe, relative to S He$\alpha$), plus the centroids of the Si He$\alpha$, S He$\alpha$, Ar He$\alpha$, Ca
He$\alpha$, and Fe K$\alpha$ blends. In the synthetic spectra, the properties of the line emission are sensitive to
variations in $\beta$, $\rho_{AM}$, and the Type Ia SN explosion model. The model with 14 Gaussian lines used in
\S~\ref{sec:Line-emission} to derive the observed values for the thirteen diagnostic quantities cannot be applied to
these synthetic spectra, because some of its underlying assumptions (like the fixed value for the He$\gamma$/He$\beta$
flux ratios) might break down in extreme cases. However, the calculation of line fluxes and centroids in the synthetic
spectra is straightforward if it is performed before convolution with an instrumental response. In this format, the
lines that contribute to a given blend can be singled out in most cases without the risk of contamination from
neighboring lines, and the continuum can be subtracted easily.  The selection energy windows for each of the lines in
the unconvolved model spectra are listed in Table \ref{tab-3}.

We have verified that these two methods used to determine line fluxes and ratios in the observed spectra and in the
synthetic spectra give consistent results. To do so, we have generated fake data from one particular model with line
emission similar to that of the Tycho SNR (from those selected in \S~\ref{sub:Successful-models}), and we have fitted
the fake data with the 14 Gaussian line model.  Since our synthetic spectra only include thermal emission, we
substituted the power law continuum by thermal bremsstrahlung in the fit. The line centroids derived
from the Gaussian fit were indistinguishable from those obtained directly from the unconvolved synthetic spectrum,
except for Ar He$\alpha$, where a $\sim0.4\%$ deviation was observed. The deviation in the line fluxes never
exceeded $7\%$, except again in the case of Ar He$\alpha$, where it was $11\%$.

Before we proceed with the comparison between models and observations, it is important to review all the possible
sources of uncertainty in the line fluxes and centroids, both for the observed spectrum and for the synthetic spectra:

\paragraph{Instrumental Limitations}
Several issues related to instrument calibration and data processing can limit the accuracy of the energies and fluxes
obtained from X-ray spectra, but these limits are often hard to estimate. For the instruments on board \textit{XMM-Newton},
\citet{cassam03:kepler} list the deviations found in the centroids of strong lines in a single observation of the Kepler
SNR.  The maximum deviations between the values measured by the different EPIC cameras (MOS1, MOS2, and PN) were of 13.5
eV (0.7\%) for Si He$\alpha$, 16.1 eV (0.7\%) for S He$\alpha$ and 28.1 eV (0.4\%) for Fe K$\alpha$ \citep[see Table 2
in][]{cassam03:kepler}. \citeauthor{cassam03:kepler} also give a value of $\sim10\%$ for the flux loss in single events,
which should be comparable to the value in the Tycho SNR. We use these deviations as estimates of the
instrumental accuracy for the \textit{XMM-Newton} spectra.

\paragraph{Interstellar Absorption}
The unabsorbed line fluxes determined from the observed spectra depend on the adopted value of the interstellar
absorption. Varying $N_{H}$ between $0.4$ and $0.8 \cdot 10^{22}\, \mathrm{cm^{-2}}$ leads to $\sim10\%$ deviations in
the Si He$\alpha$ flux with respect to the value obtained with our fiducial hydrogen column density of $0.6 \cdot
10^{22}\, \mathrm{cm^{-2}}$. For the S He$\alpha$ blend, the deviations are $\sim5\%$, and they become smaller for lines
above 3 keV. The Si He$\alpha$/S He$\alpha$ flux ratio, however, is never affected by more than $\sim5\%$, because the
flux variations are always correlated.

\paragraph{Line Extraction and Overlap}
Due to the complexity of the line emission in the thermal spectra of SNRs, it can be difficult to isolate the
contribution of a single line. In the observed spectrum, the Ar He$\alpha$ and Ca He$\alpha$ blends are most affected by
this due to the presence of neighboring weak lines (e.g., S He$\gamma$ and Ar He$\gamma$) that cannot be measured
directly\footnote{This might be the origin of the small discrepancies in the centroid and flux of Ar He$\alpha$
  mentioned above between the Gaussian line fit to the fake data and the windowing of the unconvolved synthetic
  spectrum.}. In the synthetic spectra, this problem can be minimized by using the extraction windows listed in Table
\ref{tab-3}. The only conflict that cannot be resolved in this way is in the Ca He$\alpha$ blend, which completely
overlaps with Ar He$\gamma$. In any case, this should not have a major impact on the derived Ca He$\alpha$ fluxes and
centroids for the synthetic spectra, because the flux in the Ar He$\gamma$ line only amounts to a few percent of the
total flux in the Ca He$\alpha$ energy window (see Table \ref{tab-1}).

\paragraph{Atomic Data}
In Papers I and II, we used the Hamilton \& Sarazin (HS) code to calculate synthetic spectra \citep[for a description of
the original code and the most important updates, see][]{hamilton83:emission_code,borkowski01:sedov}. This code is
implemented in XSPEC, and several simple NEI spectral models in this software package make use of its atomic data by
default (this is known as version 1.1 of the NEI atomic data, or NEI v1.1 for short). However, the HS code is not
adequate to model the high quality \textit{XMM-Newton} spectrum of Tycho, and we used the atomic data from version 2.0
of the NEI models in XSPEC instead. These atomic data are from the ATOMDB data base available at
http://cxc.harvard.edu/atomdb/ \citep[for a description, see][]{smith01:H_like_and_He_like_ions}. We augmented this
atomic data base by adding inner-shell processes, which are missing in NEI v2.0, but are important for transient
plasmas. We used the most recent published atomic calculations for inner-shell processes, such as the K-shell electron
impact excitation rates or the radiative and Auger rates for K-vacancy states for $\mathrm{Fe^{+16}}$ to
$\mathrm{Fe^{+22}}$ from \citet{bautista04:kshell} and \citet{palmeri03:fe17-24}.  Because the relevant ionization and
excitation cross sections for many ions of interest are not available, extrapolation along electronic isosequences was
often necessary.  The quality of inner-shell atomic data varies greatly among ions, resulting in potentially large
errors that are hard to estimate at this time. There are also other problems with the atomic data from ATOMDB in NEI
v2.0. This data base was originally designed for plasmas in collisional ionization equilibrium, and sometimes line
emissivities are not reliable under extreme NEI conditions. For example, we find significant differences in the L-shell
emission from Fe between NEI v1.1 and v2.0 under conditions relevant for the shocked ejecta in the Tycho SNR, although
both codes are based on the atomic calculations of \citet{liedahl95:Fe_L_lines} and should produce similar results. This
is of no consequence for the high energy line emission discussed here, but it will become important when the entire
spectrum is considered in \S~\ref{sec:Modeling-Integrated_Spectrum}. In the particular case of Fe L, we use the atomic
data from NEI v1.1.

\paragraph{Doppler Effect}
The Doppler effect has not been taken into account in the generation of the synthetic spectra. Significant Doppler
shifts affecting the entire SNR are not expected from the low bulk velocities ($>-7\cdot10^{6}\,\mathrm{cm \cdot
  s^{-1}}$) in the receding environment of Tycho determined by \citet{lee04:envir_Tycho}. Regarding shifts associated
with the dynamics of the SNR, they should be minimal in our spatially integrated spectra, which cover very large regions
of the SNR. It is possible that some of the lines in the \textit{XMM-Newton} spectrum are Doppler broadened to some
extent (see \S~\ref{sub:case-model-ddtc}), but this kind of effect is very hard to study in CCD data. We shall not
pursue this line of research in the present work.

From the preceding discussions, it is clear that a perfect agreement between models and observations is not to be
expected. In general, we have found that the statistical and instrumental uncertainties that affect the observational
data are rather small (a few times 10\% in the fluxes and below 1\% in the energies), but the systematic uncertainties
that affect the models can be much larger, and in many cases they are impossible to quantify. In this context, the
tolerance ranges on the observed diagnostic quantities should be at once generous enough to provide some flexibility in
the comparisons and restrictive enough to discriminate the most successful models. For the line flux ratios, we have
settled on a factor two range above and below the observed value.  An exception was made for the Si Ly$\alpha$/Si
He$\alpha$ and S Ly$\alpha$/S He$\alpha$ ratios, where only upper limits can be confidently derived from the
observations. For the line centroids, we imposed a tolerance range of 2\% around the observed value. In the case of Fe
K$\alpha$, which is an isolated line with well known atomic data and has the centroid at the highest energy, we imposed
a more stringent requirement of 1\% around the observed value.  These tolerance ranges are rather crude, but they are
sufficent for our present goal of distinguishing between different explosion models.

\subsection{Results for the line emission} \label{sub:Successful-models}

\begin{figure}

  \centering
  \includegraphics[scale=0.55]{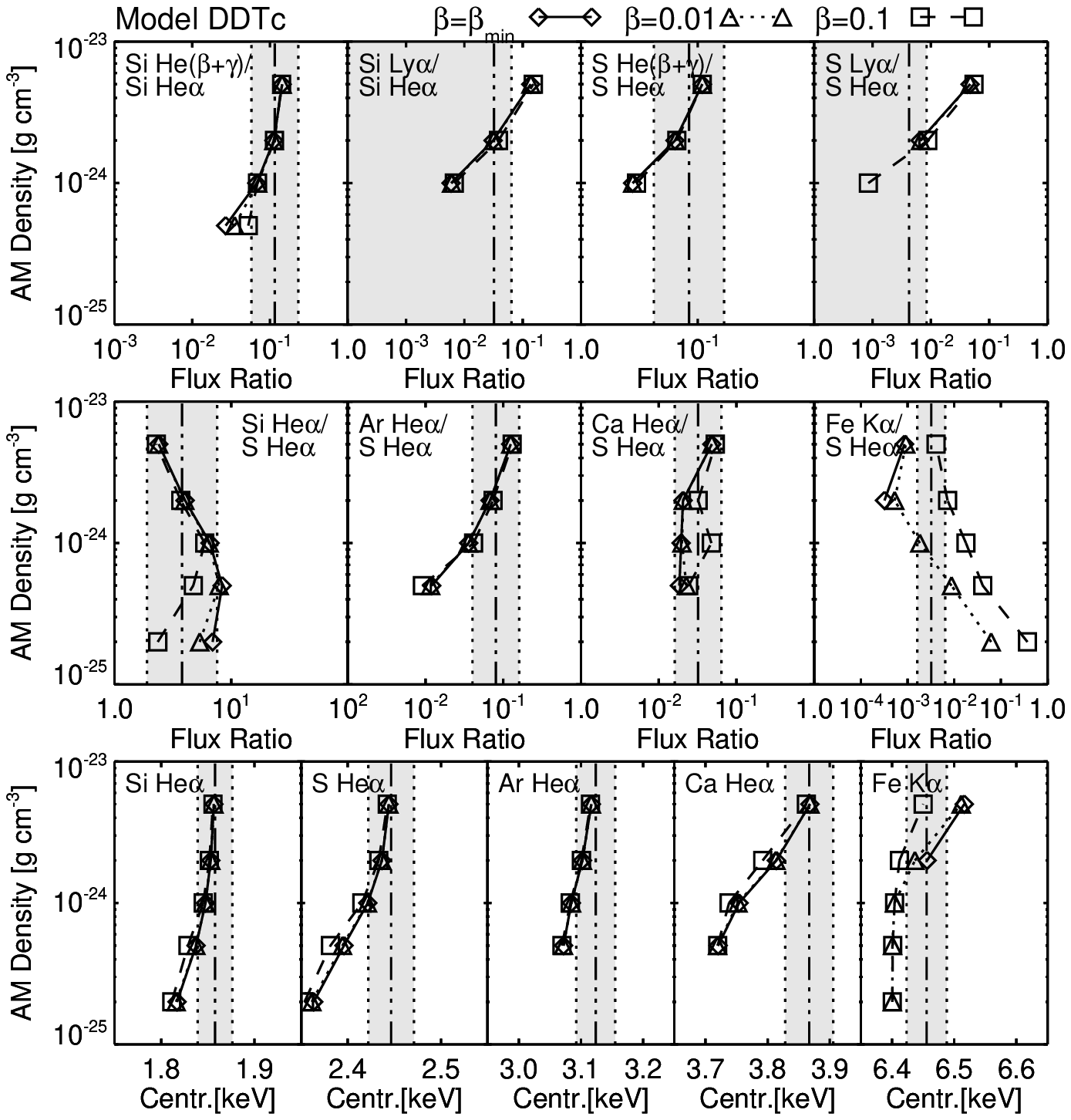}
  \includegraphics[scale=0.55]{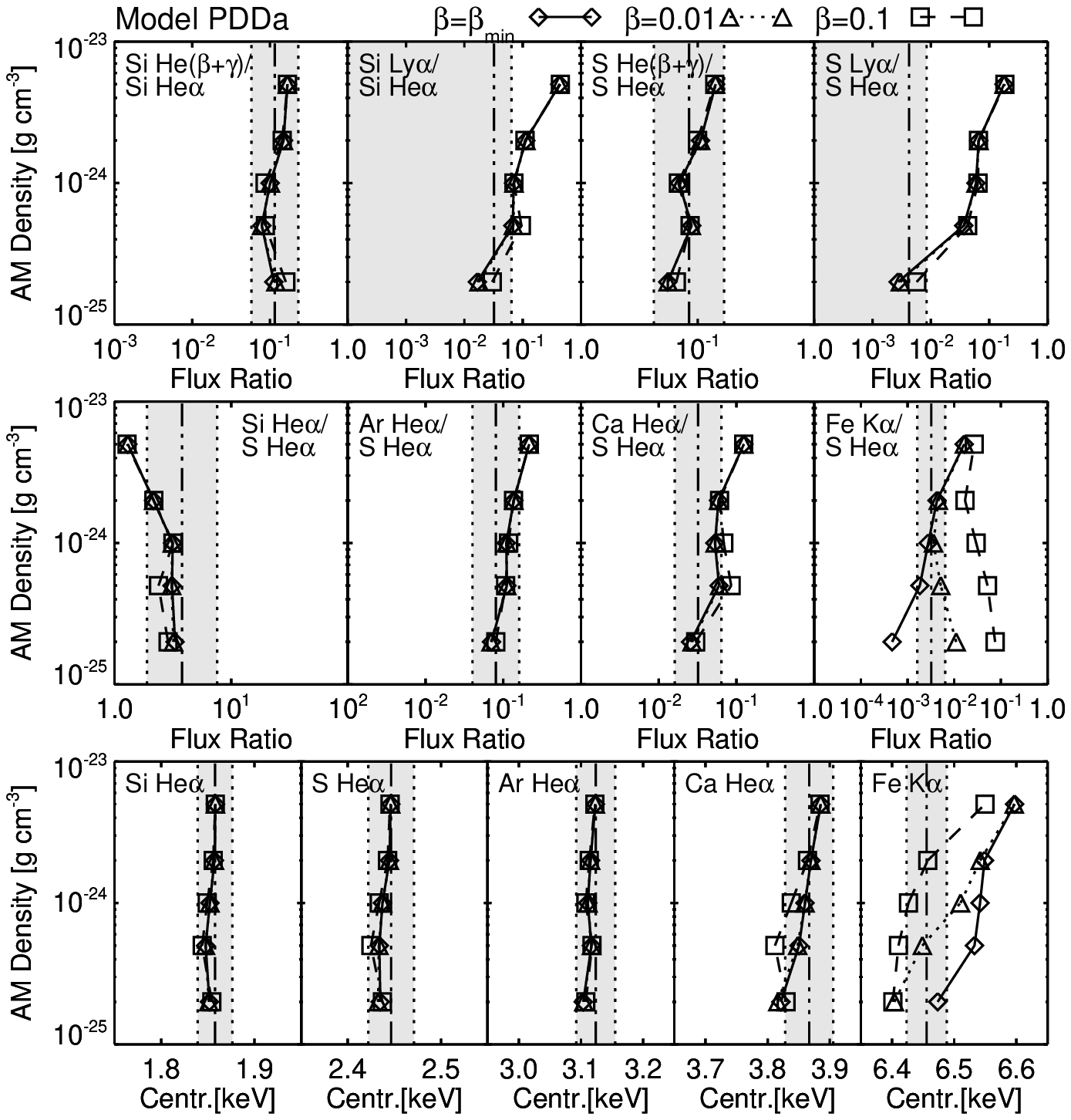}

  \caption{Comparison between the predicted line flux ratios and centroids and the observed values for models DDTc (left) and
    PDDa (right). In each case, the thirteen panels correspond to the eight diagnostic line flux ratios (top two rows)
    and the five diagnostic line centroids (bottom row). The predicted values are represented in the horizontal axis as
    a function of $\rho_{AM}$, which is mapped on the vertical axis. The values of $\beta$ are plotted with different
    styles: solid line and diamonds for $\beta_{min}$, dotted line and triangles for $\beta=0.01$, and dashed line and
    squares for $\beta=0.1$. The observed values are represented by vertical dash-triple-dotted lines, with the borders
    of the tolerance ranges described in \S~\ref{sub:comparing} as dotted lines and the `allowed' regions shaded in
    gray. Absent points denote lines that are too weak to derive a flux or a centroid. \label{fig-4}}

\end{figure}

\begin{figure}

  \centering
 
  \includegraphics[scale=0.55]{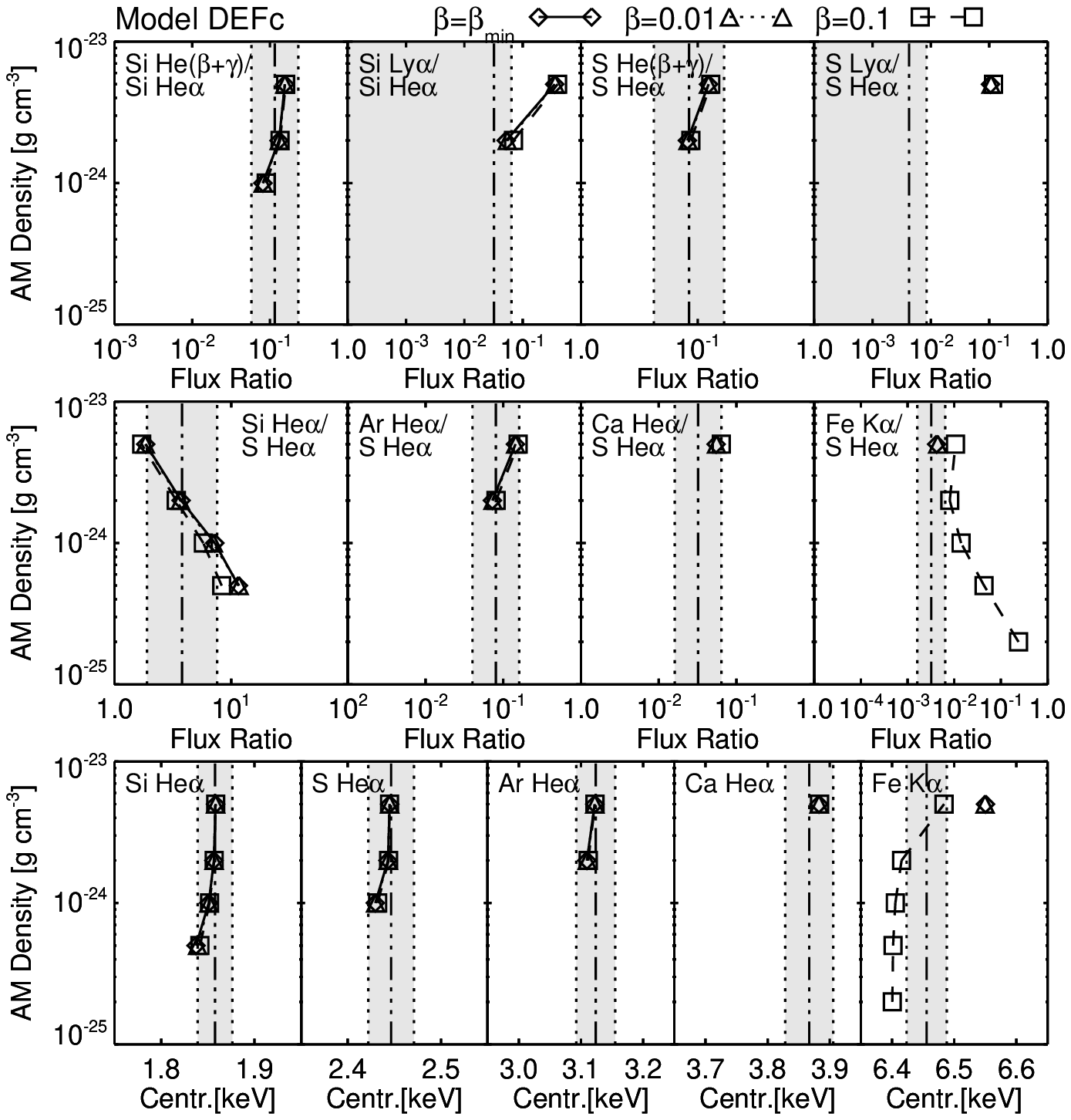}
  \includegraphics[scale=0.55]{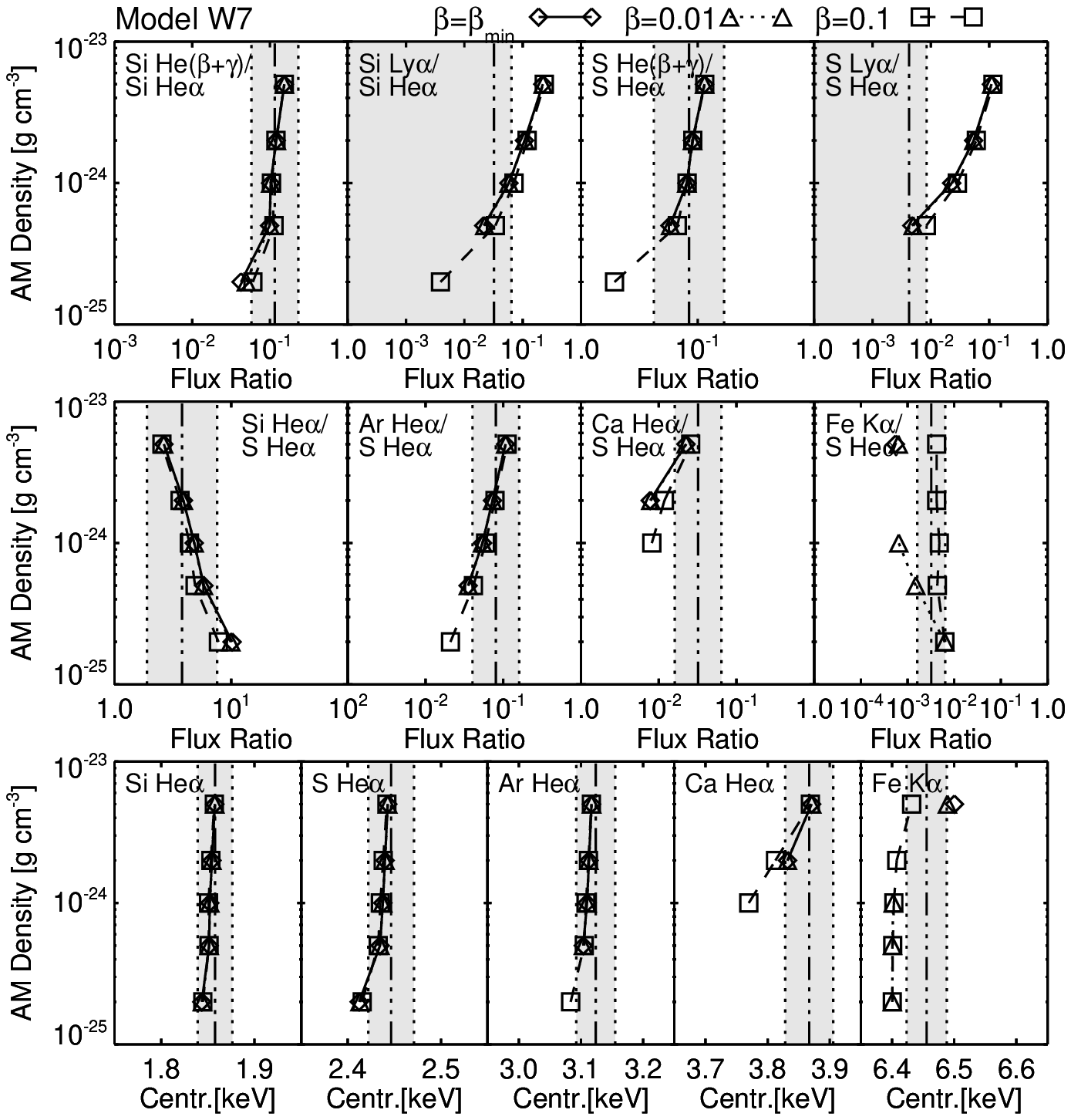}

  \caption{Comparison between the predicted line flux ratios and centroids and the observed values for models DEFc (left)
    and W7 (right). See Figure \ref{fig-4} for an explanation of the plots and labels. \label{fig-5}}

\end{figure}

\begin{figure}

  \centering
 
  \includegraphics[scale=0.55]{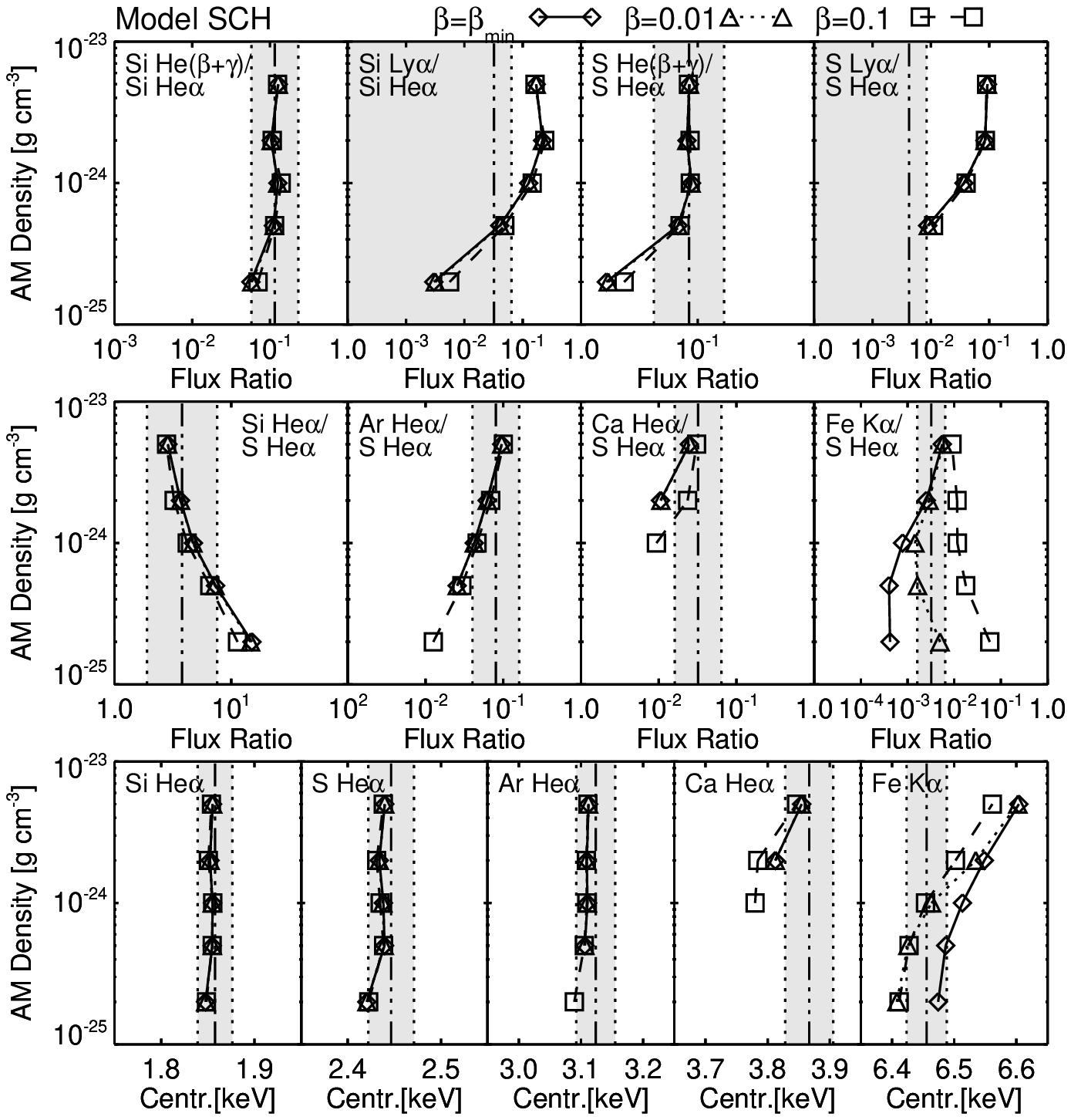}
  \includegraphics[scale=0.55]{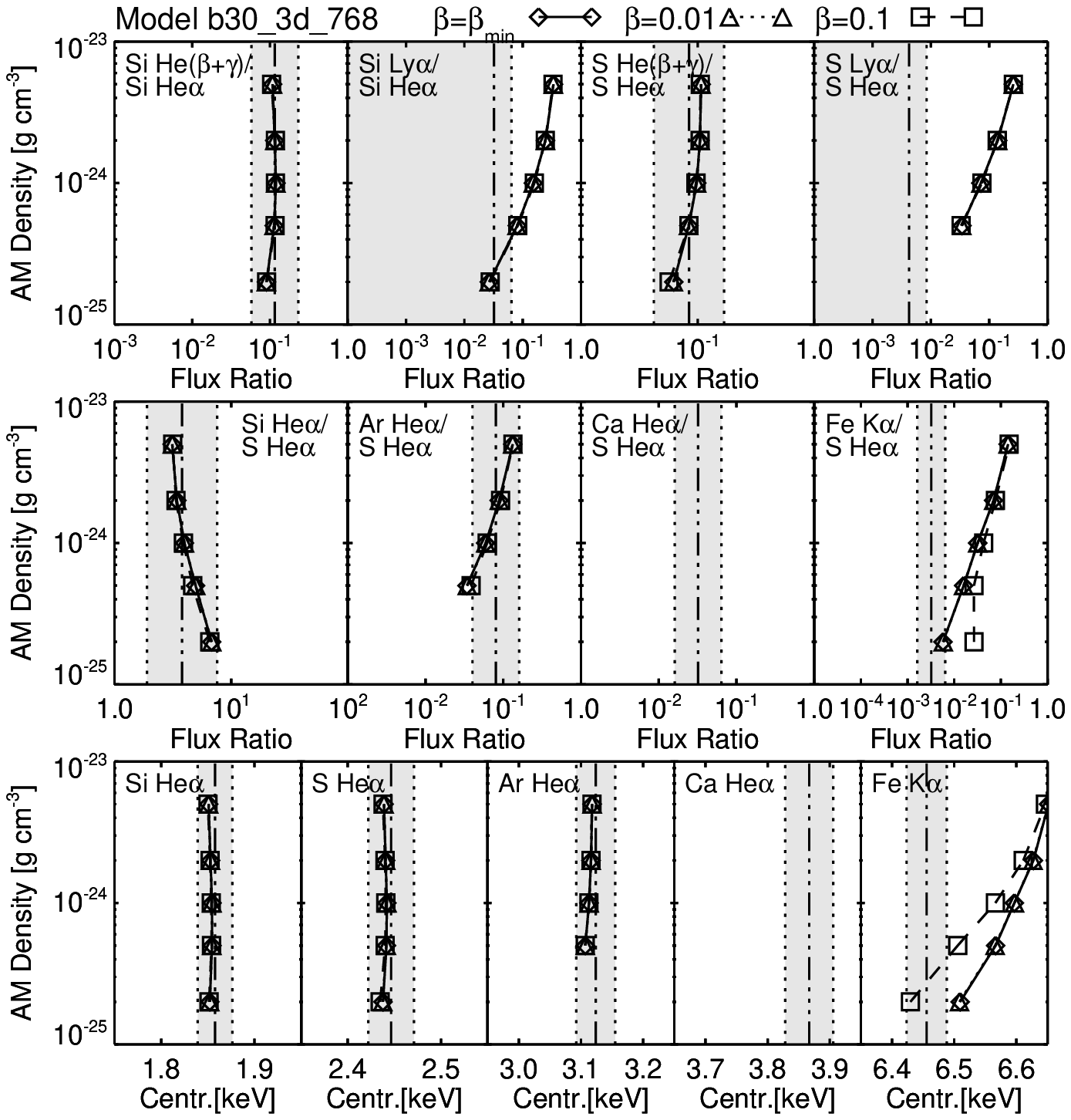}

  \caption{Comparison between the predicted line flux ratios and centroids and the observed values for models SCH (left)
    and b30\_3d\_768 (right). See Figure \ref{fig-4} for an explanation of the plots and labels. \label{fig-6}}

\end{figure}

It is not necessary to present here a complete comparison between the values of the diagnostic line flux ratios and
centroids in the 225 synthetic ejecta spectra and the observed values. Instead, we concentrate on six Type Ia SN
explosion models (DDTc, PDDa, DEFc, W7, SCH, and b30\_3d\_768) that are representative of all the pertinent features,
and we display their behavior across the $\rho_{AM}$, $\beta$ parameter space in Figures \ref{fig-4}, \ref{fig-5}, and
\ref{fig-6}. In these plots, it is easy to identify which models are able to reproduce most of the diagnostic quantities
within the specified tolerance ranges (areas shaded in gray) for a given combination of $\rho_{AM}$ and $\beta$. We have
repeated this process with each of the 15 Type Ia SN explosion models listed in \S~\ref{sub:preliminary}, identifying the
best values of $\rho_{AM}$ and $\beta$ and noting how many of the diagnostic quantities are reproduced - the results are
given in Tables \ref{tab-4} and \ref{tab-5}.

Before discussing each class of models in detail, we make a few general comments about the results seen in Figures
\ref{fig-4}, \ref{fig-5}, and \ref{fig-6}. First, the choice of values for $\rho_{AM}$ and $\beta$ seems adequate in
that reasonable results are obtained in the middle range, if at all, while the extrema can be discarded in most cases.
Variations in $\rho_{AM}$ affect the line emission of all elements in most models, while variations in $\beta$ affect
primarily the Fe K$\alpha$ emission (see discussion in \S~2.4 of Paper II). As $\rho_{AM}$ increases, so does the
emission measure averaged ionization timescale $\langle n_{e}t \rangle$ in the plasma. This moves most line centroids
towards higher energies, and increases line fluxes. In particular, the Ly$\alpha$/He$\alpha$ ratios of Si and S pose
very important constraints on the maximum allowed $\rho_{AM}$. Further constraints come from the Ca He$\alpha$ centroid,
which is displaced towards lower energies at low $\rho_{AM}$ in most models, and from the Fe K$\alpha$ centroid, which
is displaced towards Fe He$\alpha$ ($\sim6.7$ keV) at high $\rho_{AM}$ in most models. An increase of $\beta$ leads to
higher electron temperatures towards the reverse shock (see \S~2.2 and \S~2.4 in Paper II).  This enhances the
emissivity of the Fe K$\alpha$ line, but also inhibits collisional ionization, displacing the Fe K$\alpha$ centroid
towards lower energies. Within these general trends, the detailed behavior of each class of models has its own
peculiarities:

\paragraph{Delayed detonations (DDT)}
This is by far the most successful class of models. In the case of model DDTc (see Fig. \ref{fig-4}), 12 of the 13
diagnostic quantities are reproduced for $\rho_{AM}=2.0 \cdot 10^{-24}\, \mathrm{g \cdot cm^{-3}}$ and $\beta$ between
0.01 and 0.1. The only discrepant quantity is the centroid of the Ca He$\alpha$ blend, whose energy is slightly
underpredicted. The other DDT models from the grid, DDTa and DDTe, have a very similar behavior, as shown in Tables
\ref{tab-4} and \ref{tab-5}. The off-grid delayed detonation 5p0z22.25 is also very successful, but it requires a higher
AM density of $5.0 \cdot 10^{-24}\, \mathrm{g \cdot cm^{-3}}$.

\paragraph{Pulsating delayed detonations (PDD)}
This class of models is also relatively successful, specially models PDDa (see Figure \ref{fig-4}) and PDDc, which
manage to reproduce 12 and 11 of the 13 diagnostic quantities, respectively. As was discussed in \S~4.2 of Paper I, PDD
models have chemical composition profiles that are very similar to DDT models, but their density profiles are much
steeper. This leads to higher values of $\langle n_{e}t \rangle$ in the shocked plasma so that the Ly$\alpha$/He$\alpha$
ratios of Si and S become incompatible with the observations for $\rho_{AM}$ above $2.0\cdot 10^{-25}\, \mathrm{g \cdot
  cm^{-3}}$. The high ionization timescales in the ejecta also affect the Fe K$\alpha$ centroid, which moves rapidly to
higher energies as Fe He$\alpha$ lines become more prominent. We note that PDD models provide a better approximation to
the Ca He$\alpha$ centroid than DDT models.

\paragraph{One-dimensional deflagrations (DEF)}
The one-dimensional deflagrations from our model grid have several serious shortcomings. As can be seen in Figure
\ref{fig-5} for model DEFc, many important lines only appear at the highest values of $\rho_{AM}$, and even then the
properties of the Si and S line emission are hard to reconcile with those of other elements. Models DEFa and DEFf behave
similarly to model DEFc, and none of them can reproduce more than 9 of the 13 diagnostic quantities at once, even under
the most favorable conditions. Models DEFc and DEFf at $\rho_{AM}=5.0 \cdot 10^{-24}\, \mathrm{g \cdot cm^{-3}}$ have
the peculiarity of being the only models where radiative losses are of some importance at the age of Tycho. In these two
models, approximately 0.05 $\mathrm{M_{\odot}}$ of C and O in the outermost ejecta might undergo runaway cooling. This
is only a minor effect, and will not modify our conclusions, because the line emission from the other elements is not
affected. 

\paragraph{One-dimensional `fast' deflagrations (W7)}
It is interesting to analyze the performance of this model, because it has been used in attempts to reproduce the X-ray
emission from the Tycho SNR by several authors
\citep{hamilton86:Tycho,itoh88:Tycho_X_rays,brinkmann89:NEI_SNRs_Tycho,sorokina04:typeIasnrs}. All these previous works
used NEI calculations coupled to hydrodynamic models with varying degrees of sophistication, and all of them came to the
same conclusion: the W7 model does not have enough Fe in the outer layers of ejecta to reproduce the Fe K$\alpha$ flux
in the Tycho SNR. Our simulations without collisionless electron heating ($\beta=\beta_{min}$) are in agreement with
this, as shown by the absence of diamonds in the corresponding panels of Fig. \ref{fig-5}. Nevertheless, we note that
partial collisionless heating of electrons at the reverse shock ($\beta=0.1$) can explain the Fe K$\alpha$/S He$\alpha$
flux ratio for a wide range of $\rho_{AM}$ values (see the squares in the same panels of Fig. \ref{fig-5}). In any case,
model W7 can only reproduce 9 of the 13 diagnostic quantities, and must be discarded.

\paragraph{Sub-Chandrasekhar explosions (SCH)}
Sub-Chandrasekhar explosions lead to a complex structure in the ejecta, with an outer region dominated by Fe-peak nuclei
bound between two strong density peaks (see \S~2 in Paper I). This results in high values of $\langle n_{e}t \rangle$,
which in turn restrict the range of $\rho_{AM}$ that can reproduce both the Fe K$\alpha$ centroid and the
Ly$\alpha$/He$\alpha$ ratio of Si and S (see Fig. \ref{fig-6}). At these low ambient medium densities, the emission
measure of Ar and Ca in the shocked ejecta is not high enough to account for the observed line emission.

\paragraph{Prompt detonations (DET)}
These models are known to be unrealistic for Type Ia SNe, because the ejecta are essentially pure Fe with only trace
amounts of intermediate mass elements. Not surprisingly, they fail to reproduce the line emission from all the elements
in the SNR spectrum except Fe. We do not show plots comparing model and observations in this case.

\paragraph{Three-dimensional deflagrations (B30U, b30\_3d\_768)}
These models represent the most sophisticated self-consistent calculations of Type Ia SN explosions that have been
performed to date. Model b30\_3d\_768 (see Figure \ref{fig-6}) is more successful in reproducing the line emission from
Tycho than model B30U due to its higher Ar and Ca content. Yet, its predictions are clearly in conflict with the
observations, mainly due to the presence of large amounts of Fe in the outer ejecta, where the plasma is hotter and more
highly ionized. Under these conditions, the centroid of the Fe K$\alpha$ blend and the Fe K$\alpha$/S He$\alpha$ ratio
do not match the observations, even for the lowest values of $\rho_{AM}$. In Paper II, we suggested that the high degree
of mixing in the ejecta that is characteristic of all three-dimensional deflagrations is in conflict with the X-ray
observations of several Type Ia SNRs, which seem to indicate that Fe and Si are emitting under different physical
conditions. This conclusion is confirmed by the more detailed work presented here.

\section{MODELING THE SPATIALLY INTEGRATED SPECTRUM} \label{sec:Modeling-Integrated_Spectrum}

\subsection{Comparing the models to the observed spectrum}

After identifying the most promising explosion models as decribed in \S~\ref{sec:Line-emission-models}, the next step is
to compare them to the spatially integrated spectrum. In order to perform these comparisons, we assume that
all the emission from the shocked AM is nonthermal, and describe it with a power law continuum of index 2.72 (see
discussion in \S~\ref{sub:Obs-X-ray}). This assumption is based on the results of \citet{warren05:Tycho}, who used the
morphology of the rim emission to constrain the contribution to the X-ray flux from a thermal shocked AM component, and
found an upper limit of 9\% in a subset of our region B ($226 \leq \theta \leq 271$). We note that this limit might be
somewhat higher in other parts of region B where the rim is thicker, but in any case the temperature of the thermal
component would have to be well below the 2 keV determined by \citet{hwang98:tycho-Feemission}, because the high energy
(E$>$1.5 keV) continuum is known to be dominated by nonthermal emission. Since the exact value of the hydrogen column
density is unknown, the properties of this low temperature AM component would be very difficult to determine from the
integrated spectrum. For simplicity, we assume that its contribution is negligible.

As we mentioned in \S~\ref{sec:Goals-and-strategy}, direct comparison between our models and the observed spectrum is
not straightforward. Given the limitations of the models and the substantial uncertainties in the atomic data included
in the spectral codes, it is unrealistic to expect a `valid fit' from the statistical point of view (i.e., with a
reduced $\chi^{2}$ below the required limit). Under these circumstances, it is not appropriate to rely solely on the
value of $\chi^{2}$ as a measure of the best `fit', and we use additional criteria to decide which models are most
satisfactory. In our case, there are only three parameters that can be adjusted for each ejecta model: the normalization
of the ejecta and AM components ($norm_{ej}$, $norm_{AM}$) and the value of the hydrogen column density $N_{H}$. We will
adjust these parameters in two steps. In a first step, the normalization of the two components will be determined by
minimizing the $\chi^{2}$ statistic in the spectrum between 1.6 and 9.0 keV, with the hydrogen column density set to our
fiducial value of $N_{H}=0.6\cdot10^{22}\,\mathrm{cm^{-2}}$. By doing this, we guarantee that the high energy spectrum
will be approximated to the best ability of the ejecta model. In a second step, the normalization parameters will be
frozen and the interstellar absorption will be adjusted, again by minimizing the $\chi^{2}$ statistic, but this time
using the spectrum between 0.8 keV (the location of the brightest line in the Fe L complex) and 10 keV. The goal of this
second adjustment is to determine whether the ejecta model is capable of reproducing the Fe L/Fe K$\alpha$ ratio for any
given value of $N_{H}$. No adjustment whatsoever will be performed on the spectrum below 0.8 keV in order to prevent
ejecta models with excess emission in this region from forcing high values of $N_{H}$ that might disrupt the Fe L/Fe
K$\alpha$ ratio.

In addition to the quality of the spectral `approximation' (as opposed to fit) provided by each ejecta model, we employ
other consistency checks to evaluate its performance. First, $N_{H}$ should not depart significantly from the values
discussed in \S~\ref{sec:Line-emission}. Any model requiring $N_{H}$ to be much higher than $0.8$ or much lower than
$0.4\cdot10^{22}\,\mathrm{cm^{-2}}$ is probably overpredicting or underpredicting the amount of Fe L emission. Second,
the normalization distance $D_{norm}$ inferred from the value of $norm_{ej}$ should be within the 1.5-3.1 kpc range
mentioned in \S~\ref{sub:Obs-radio-optical}. The value of $D_{norm}$ can be determined using the equation
\begin{equation}
  D_{norm}=\frac{10 \mathrm{kpc}}{\sqrt{\xi \cdot norm_{ej}}}
\end{equation}
where 10 kpc is the fiducial distance to the source used in the calculation of the ejecta models and $\xi$ is a
correction factor that accounts for the incomplete spatial coverage of region B (which only contains photons from $\sim
30\%$ of the SNR surface, or $1/\xi$ of the total flux). A more accurate value of $\xi$ can be found by comparing the
fitted line fluxes for the brightest lines in region B listed in Table \ref{tab-1} to the fluxes for the entire SNR
determined by \citet{hwang97:tycho-ASCA}. For the four brightest line blends, this flux ratio is equal to 2.91 (Si
He$\alpha$), 2.77 (Si He$\beta$), 2.87 (S He$\beta$), and 2.82 (Ar He$\alpha$); we adopt a value of 2.8 for $\xi$ as the
best approximation to the mean.  Third, the value of $norm_{AM}$ should make the power law the main contributor to the
continuum at high energies, in agreement with the results of \citet{warren05:Tycho}. Quantitatively, $norm_{AM}$
corresponds to the the normalized flux at 1 keV from the AM in region B, expressed in units of $\mathrm{phot \cdot
  cm^{-2} \cdot s^{-1} \cdot keV^{-1}}$.  This number can be converted to a flux from the shocked AM in the entire
remnant using the correction factor $\xi$, so that $F_{AM} = \xi \cdot norm_{AM}$ should be within the $7.4-8.9\,
\mathrm{phot \cdot cm^{-2} \cdot s^{-1} \cdot keV^{-1}}$ range determined by \citet{fink94:tycho-ginga}. Finally, an
independent constraint on the distance is provided by matching the angular sizes of the fluid discontinuities (RS, CD
and FS) determined by \citet{warren05:Tycho}. Since the dynamics of the FS are strongly modified by CR acceleration and
the surface of the CD is corrugated due to dynamic instabilities, the only straightforward match that we can perform to
our one-dimensional models with $\gamma=5/3$ is the angular radius of the RS. We denote by $D_{RS}$ the distance
required by each model to reproduce the 183'' RS radius found by \citet{warren05:Tycho}.

\subsection{Results for the spatially integrated spectra} \label{sub:performance-integrated}

\begin{figure}

  \centering
 
  \includegraphics[scale=0.55]{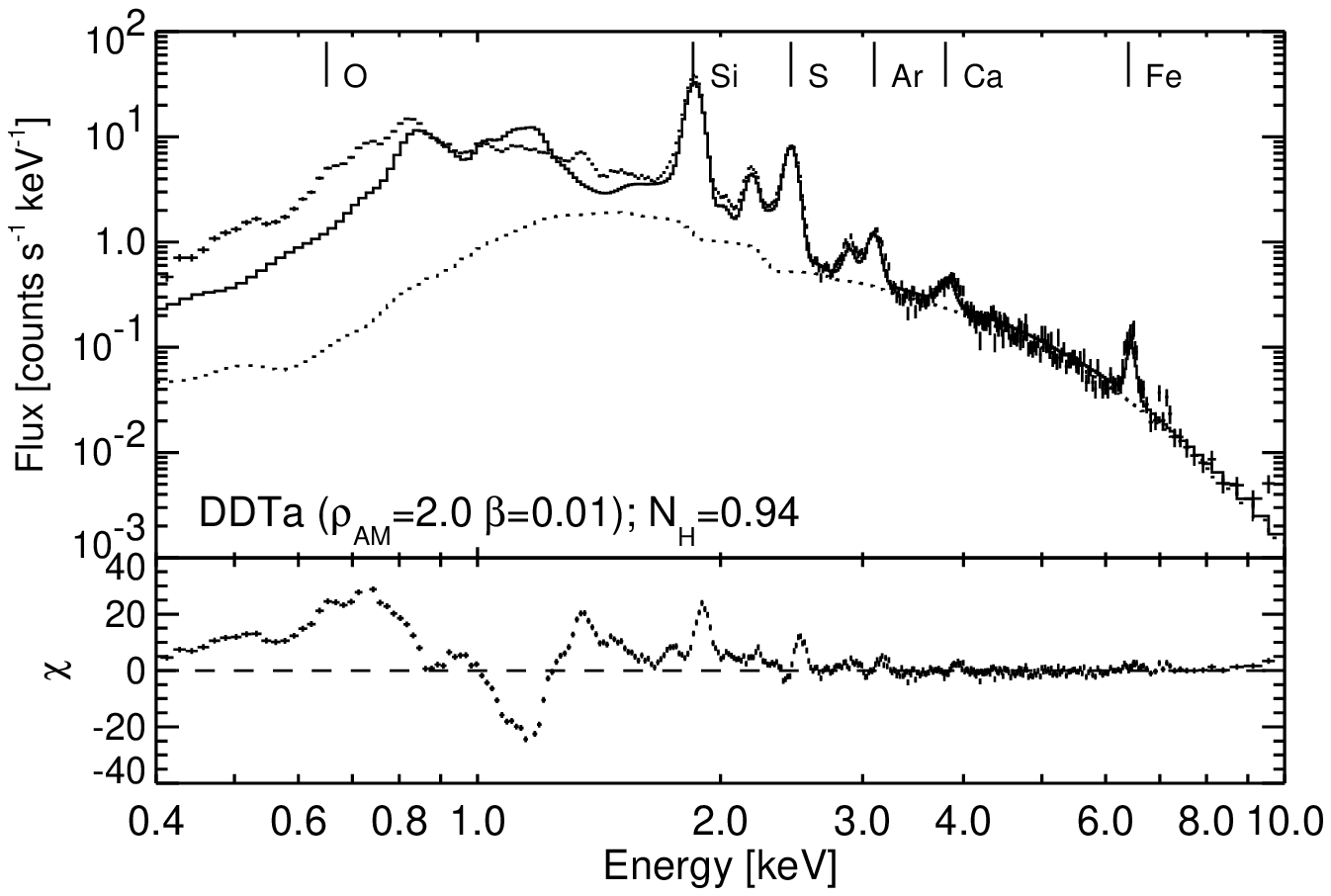}
  \includegraphics[scale=0.55]{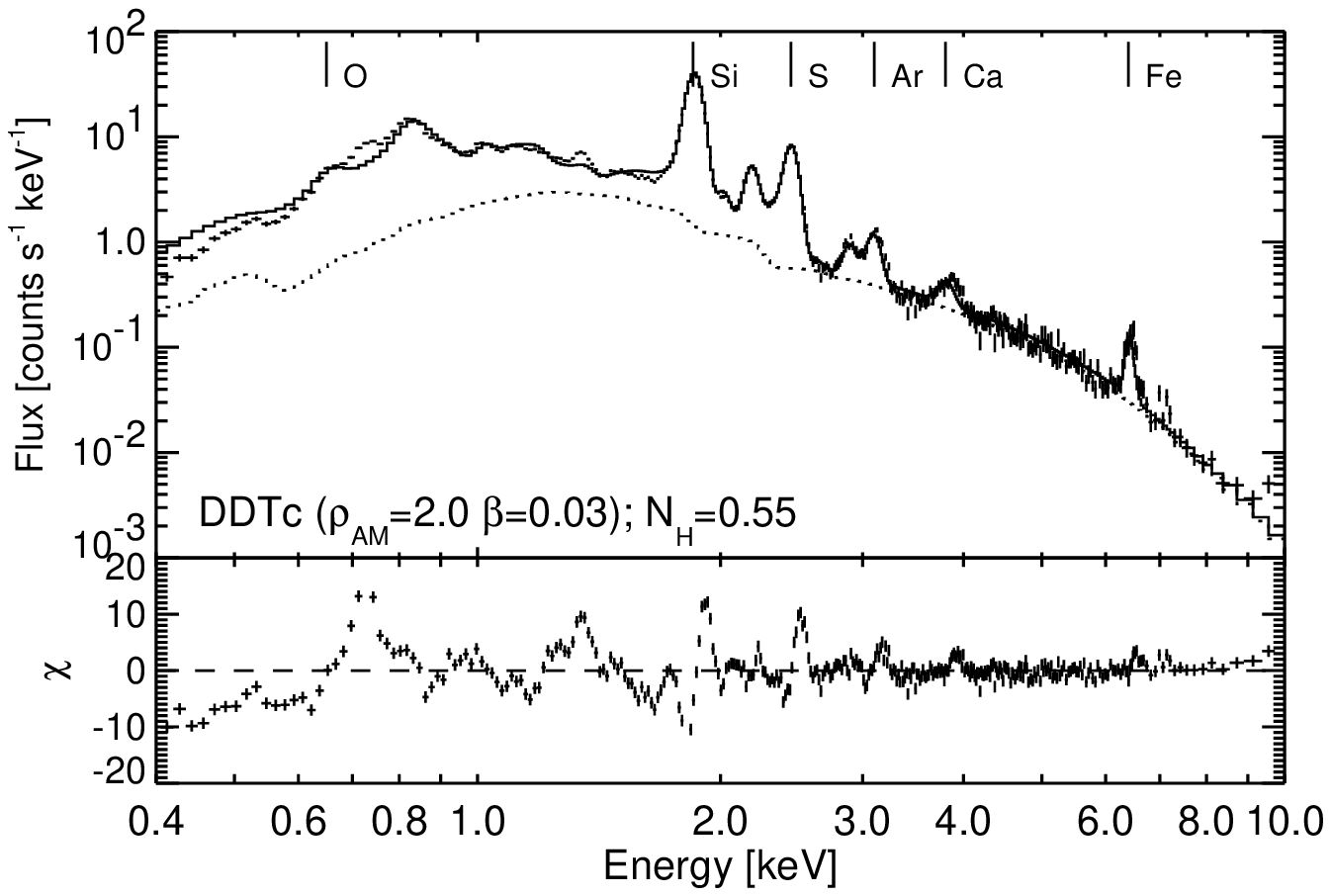}
  \includegraphics[scale=0.55]{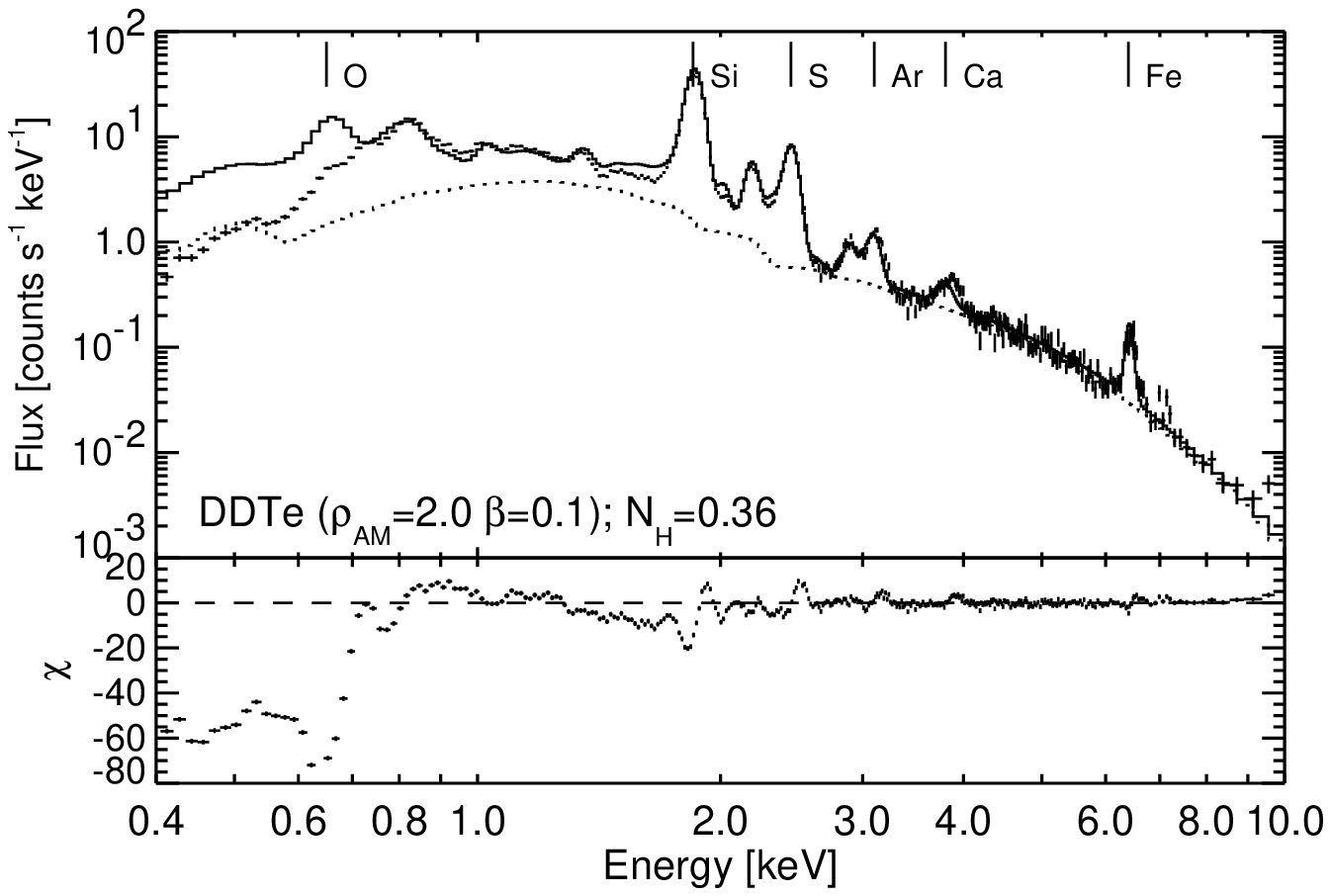}
  \includegraphics[scale=0.55]{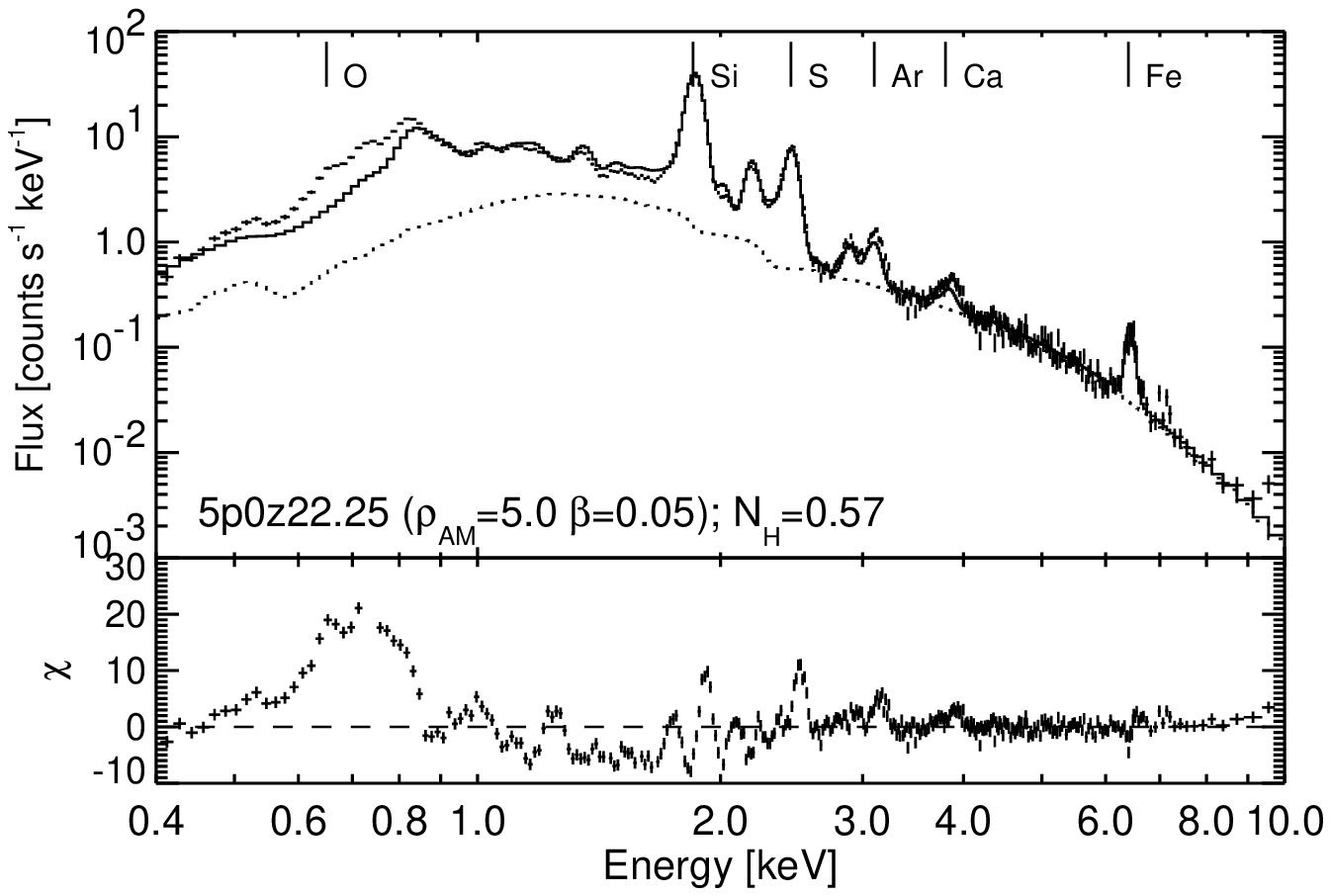}
  \includegraphics[scale=0.55]{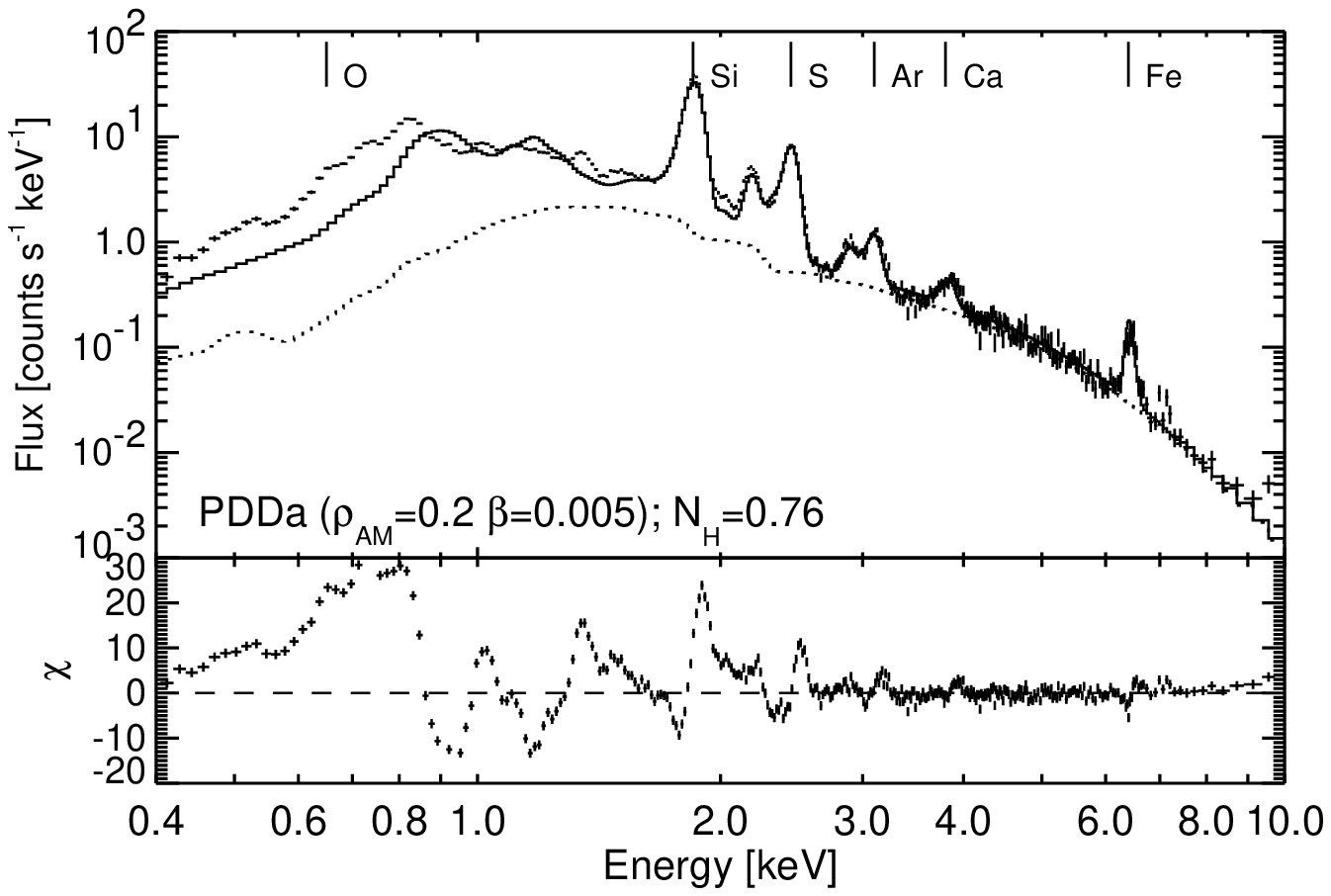}
  \includegraphics[scale=0.55]{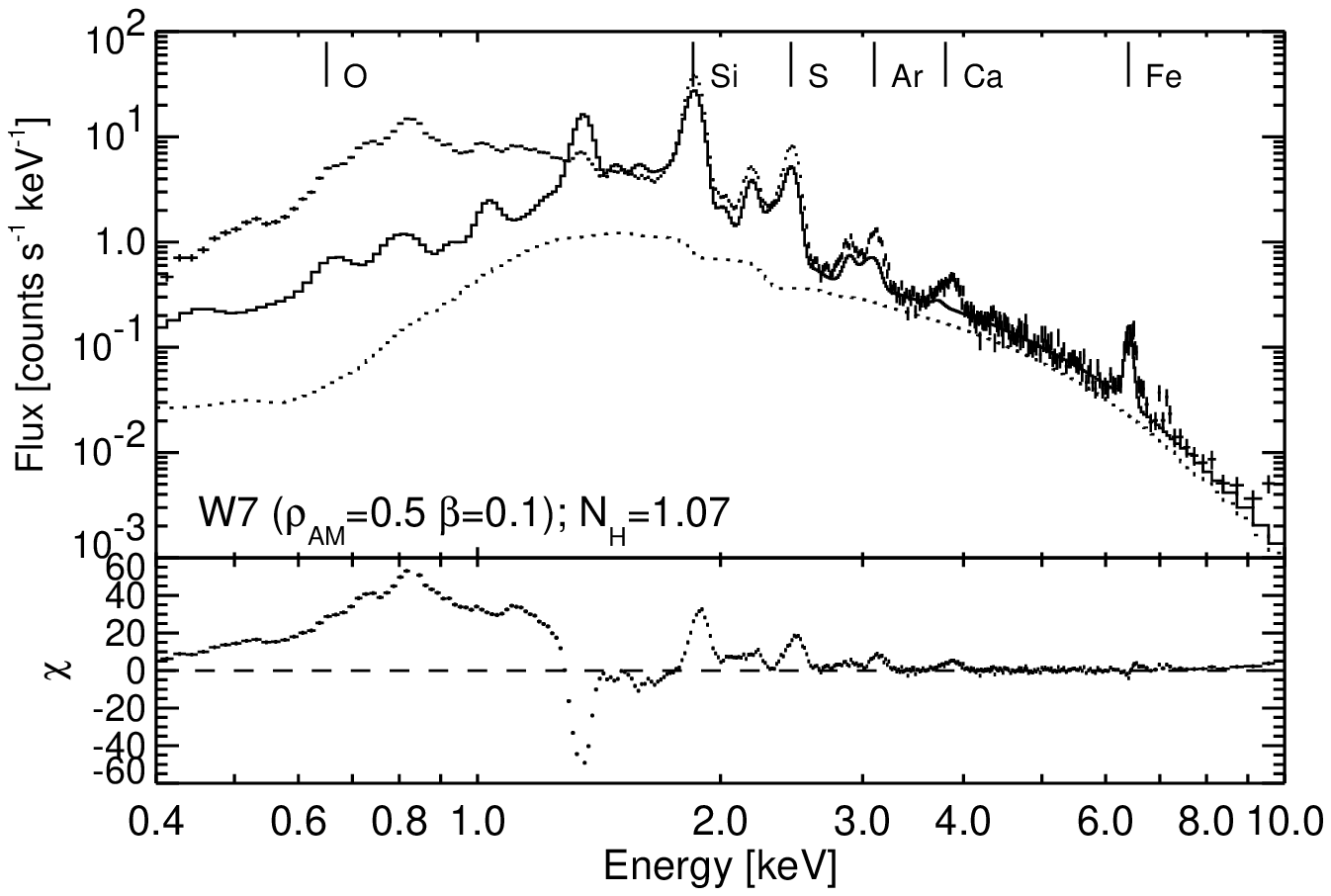}
  \includegraphics[scale=0.55]{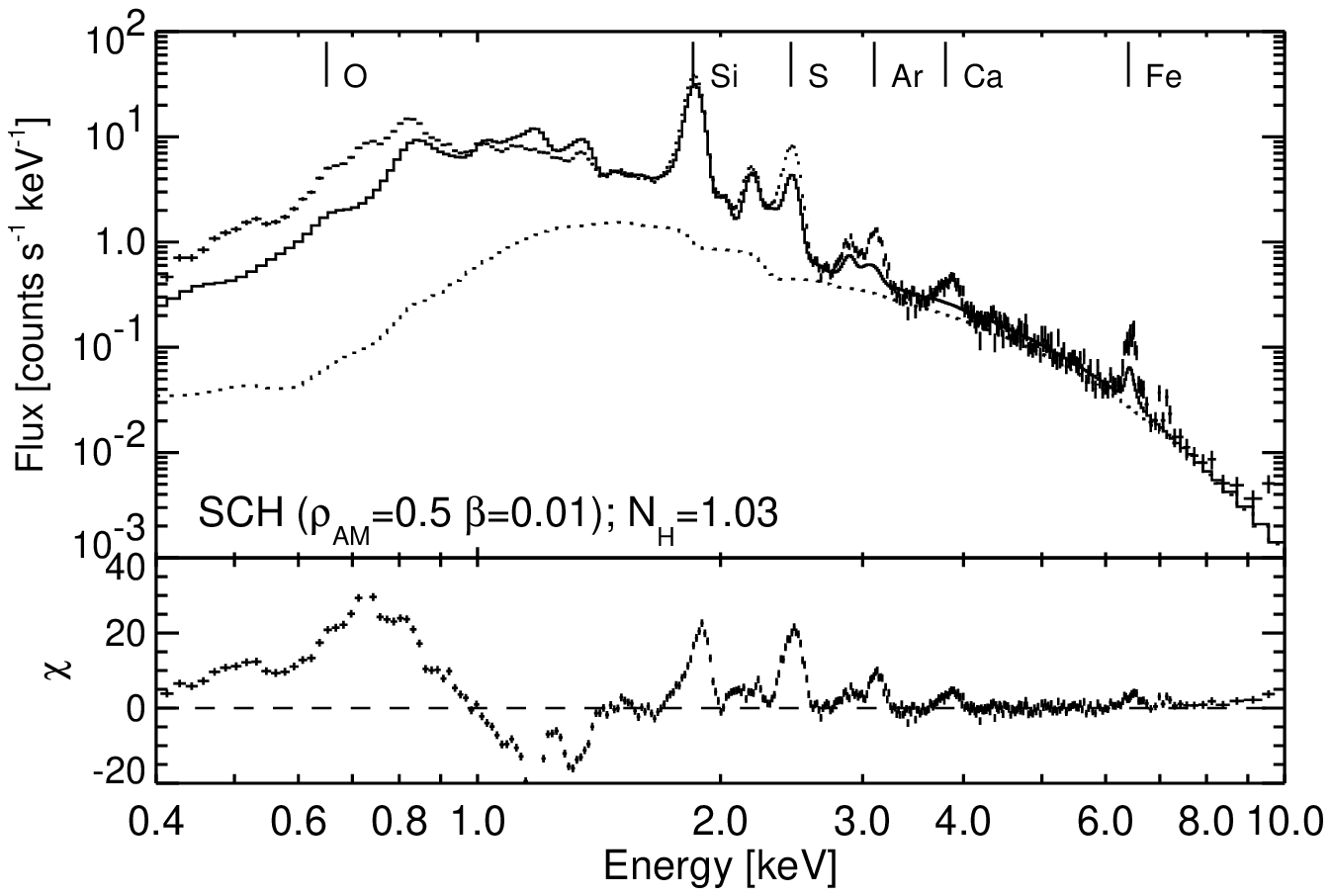}
  \includegraphics[scale=0.55]{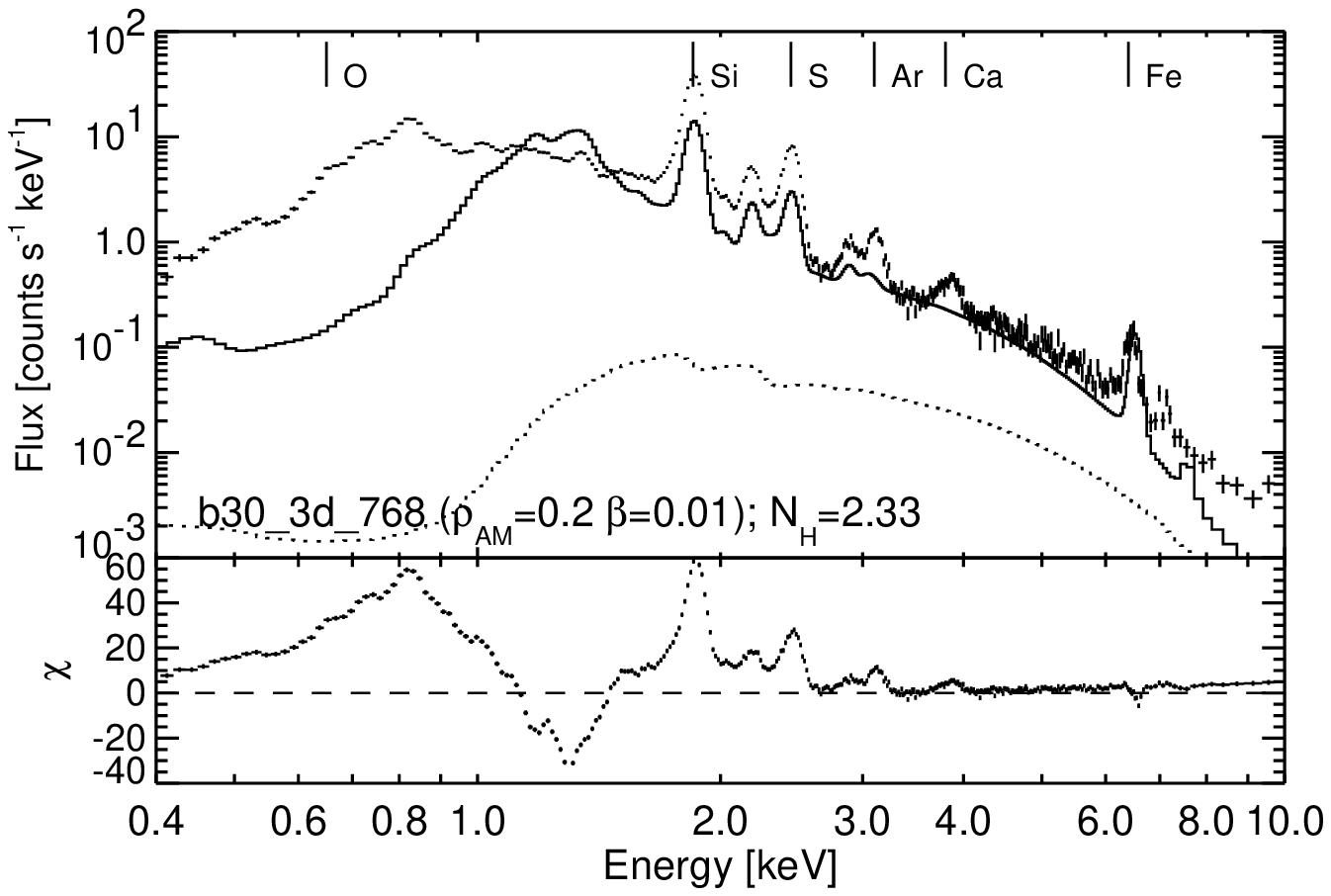}

  \caption{Comparison between the ejecta models and the spatially integrated spectrum of region B. The ejecta model, with the
    corresponding values $\rho_{AM}$ (in units of $10^{-24}\,\mathrm{g \cdot cm^{-3}}$) and $\beta$, is indicated in
    each plot, as well as the adjusted value of $N_{H}$. The power law component is displayed for clarity (dotted line).
    \label{fig-7}}

\end{figure}

In Figure \ref{fig-7}, we compare the spatially integrated spectrum from region B with eight ejecta models selected from
our grid. Four of these models have proven the most successful in reproducing the diagnostic quantities for the line
emission (see \S~\ref{sub:Successful-models}): DDTa ($\rho_{AM}=2 \cdot 10^{-24} \, \mathrm{g \cdot cm^{-3}}$,
$\beta=0.01$), DDTc ($\rho_{AM}=2 \cdot 10^{-24} \, \mathrm{g \cdot cm^{-3}}$, $\beta=0.03$), 5p0z22.25 ($\rho_{AM}=5
\cdot 10^{-24} \, \mathrm{g \cdot cm^{-3}}$, $\beta=0.05$), and PDDa ($\rho_{AM}=2 \cdot 10^{-25} \, \mathrm{g \cdot
  cm^{-3}}$, $\beta=0.005$). Models DDTc, 5p0z22.25, and PDDa require values of $\beta$ that are off the nodes of the
original $(\rho_{AM},\beta)$ grid (see Tables \ref{tab-4} and \ref{tab-5}).  We have selected the values that provide
the best possible approximation to the Fe K$\alpha$ emission (i.e., match the observed Fe K$\alpha$/S He$\alpha$ ratio
and Fe K$\alpha$ centroid).  Four other ejecta models are also included in Figure \ref{fig-7}, mainly for illustrative
purposes. Model DDTe ($\rho_{AM}=2 \cdot 10^{-24} \, \mathrm{g \cdot cm^{-3}}$, $\beta=0.1$) completes the sequence of
DDT models, and models W7 ($\rho_{AM}=5 \cdot 10^{-25} \, \mathrm{g \cdot cm^{-3}}$, $\beta=0.1$), SCH ($\rho_{AM}=5
\cdot 10^{-25} \, \mathrm{g \cdot cm^{-3}}$, $\beta=0.01$), and b30\_3d\_768 ($\rho_{AM}=2 \cdot 10^{-25} \, \mathrm{g
  \cdot cm^{-3}}$, $\beta=0.01$) are provided to assess the viability of one-dimensional deflagrations,
sub-Chandrasekhar explosions and three-dimensional deflagrations.

The adjusted values for $N_{H}$, $norm_{AM}$, and $norm_{ej}$ are listed in Table \ref{tab-6}, along with $F_{AM}$ and
the distance estimates $D_{norm}$ and $D_{RS}$. These two distance estimates are mutually inconsistent in all the models
presented here. In general, models with higher $\rho_{AM}$ have higher fluxes in the shocked ejecta and lower radii of
the RS, which result in higher values of $D_{norm}$ and lower values of $D_{RS}$, and vice versa (the extreme examples
are models 5p0z22.25 and PDDa). We will discuss the inconsistency in the distance estimates in
\S~\ref{sub:limitations-models}.

\paragraph{Delayed detonations (DDTa, DDTc, DDTe, 5p0z22.25)}
Delayed detonation models are clearly superior to the other classes of Type Ia SN explosions. The low energy X-ray
emission poses strong constraints on the main parameter in one-dimensional DDT models, the deflagration-to-detonation
transition density $\rho_{tr}$. Model DDTa ($\rho_{AM}=2 \cdot 10^{-24} \, \mathrm{g \cdot cm^{-3}}$, $\beta=0.01$),
which has the highest $\rho_{tr}$ and therefore synthesizes more Fe and less intermediate mass elements than the other
models, overpredicts the flux in the Fe L complex, requiring a hydrogen column density of $0.94
\cdot10^{22}\,\mathrm{cm^{-2}}$ to compensate this. The shocked Fe is also overionized, as indicated by the excess flux
coming from the L-shell lines of Fe$^{+20}$ and other ions of higher charge around 1.1 keV. Model DDTe ($\rho_{AM}=2
\cdot 10^{-24} \, \mathrm{g \cdot cm^{-3}}$, $\beta=0.1$), on the other hand, is probably underpredicting the Fe L flux
judging from the low $N_{H}$ that it requires, $0.36 \cdot10^{22}\,\mathrm{cm^{-2}}$. Although the shape of the Fe L
complex is approximately correct in this case, the model clearly overpredicts the O Ly$\alpha$ flux at 0.65 keV,
indicating that there is too much O in the outer layers of the ejecta, and that $\rho_{tr}$ is probably too low.  The
best results are obtained for the intermediate model DDTc ($\rho_{AM}=2 \cdot 10^{-24} \, \mathrm{g \cdot cm^{-3}}$,
$\beta=0.03$), which can reproduce with reasonable accuracy both the shape of the Fe L complex and the flux of the O
Ly$\alpha$ line for $N_{H} = 0.55 \cdot10^{22}\,\mathrm{cm^{-2}}$. The only features that this model does not
approximate well are the Mg He$\alpha$ flux at 1.35 keV and an Fe L-shell line at 0.7 keV. A relationship between
$\rho_{tr}$ and the properties of the synthetic spectra was already hinted at by Tables \ref{tab-4} and \ref{tab-5},
because DDT and PDD models with lower $\rho_{tr}$ required lower values of $\beta$ \citep[see][for a more general
discussion of this topic]{badenes05:model_grid}. The `off-grid' model 5p0z22.25 ($\rho_{AM}=5 \cdot 10^{-24} \,
\mathrm{g \cdot cm^{-3}}$, $\beta=0.05$) is better than DDTa or DDTe, but inferior to DDTc. It requires a hydrogen
column density of $0.57 \cdot10^{22}\,\mathrm{cm^{-2}}$, and provides a good approximation to the spatially integrated
spectrum, fitting the Mg He$\alpha$ line better than DDTc, but it fails to reproduce the O Ly$\alpha$ flux. We note that
in all the DDT models presented here, the high energy continuum is dominated by the nonthermal AM component, with values
of $F_{AM}$ between $8.3$ and $7.9\cdot10^{-2} \, \mathrm{phot \cdot cm^{-2} \cdot s^{-1} \cdot keV^{-1}}$, within the
range determined by \citet{fink94:tycho-ginga}.  The distance $D_{RS}$, obtained by matching the location of the RS is
within the 1.5-3.1 kpc range in all cases, but the normalization distance $D_{norm}$ is always too large.

\paragraph{Pulsating delayed detonations (PDDa)}
The pulsating delayed detonation model PDDa ($\rho_{AM}=2 \cdot10^{-25} \, \mathrm{g \cdot cm^{-3}}$, $\beta=0.005$)
provides a good approximation to the spectrum at high energies, but fails at low energies. Although the model requires a
reasonable hydrogen column density of $0.76 \cdot10^{22}\,\mathrm{cm^{-2}}$, the prominent Fe$^{+16}$ line at $\sim0.8$ keV
does not appear in the synthetic spectrum, and neither does the O Ly$\alpha$ line. The value of $F_{AM}$ for this model,
$7.9\cdot10^{-2} \, \mathrm{phot \cdot cm^{-2} \cdot s^{-1} \cdot keV^{-1}}$, is similar to those required by the DDT
models, and also within the expected range. The low value of $\rho_{AM}$, however, implies a low emitted flux in the
shocked ejecta and a large RS radius, which result in an unrealistically low $D_{norm}$ (0.72 kpc) and a high $D_{RS}$
(4.38 kpc).

\paragraph{One-dimensional deflagrations (W7)}
In addition to the obvious shortcomings at high energies, the W7 ($\rho_{AM}=5 \cdot 10^{-25} \, \mathrm{g \cdot
  cm^{-3}}$, $\beta=0.1$) model is extremely poor at low energies. This is mainly due to the high mass of Mg in the
ejecta of W7 ($8.5\cdot10^{-3}\,M_{\odot}$, compared to $< 10^{-3}\,M_{\odot}$ in other models). In this case, it is the
high flux in the Mg He$\alpha$ complex, and not an excess of Fe L emission, that results in the high value of $N_{H}$.
The thermal continuum from the W7 model is higher than in the DDT and PDD models because of the high amount of unburned
C and O in the outer layers of ejecta, which leads to a value of $F_{AM}$ below the expected range. The distance
$D_{norm}$ is within the expectations, but $D_{RS}$ is too high.

\paragraph{Sub-Chandrasekhar Explosions (SCH)}
The SCH ($\rho_{AM}=5 \cdot 10^{-25} \, \mathrm{g \cdot cm^{-3}}$, $\beta=0.01$) model is clearly unable to reproduce
the observed Fe L/Fe K$\alpha$ flux ratio. The Fe K$\alpha$ flux can be increased by increasing $\beta$, but this moves
the centroid of this line blend to too high energies (see Fig. \ref{fig-6}). We have verified that an increase in
$\beta$ cannot solve problems with Fe L emission either: the flux is too high (as indicated by the required $N_{H}$ of
$1.03 \cdot10^{22}\,\mathrm{cm^{-2}}$) and the overall shape does not correspond with the observed spectrum. The AM flux
is $7.3\cdot10^{-2} \, \mathrm{phot \cdot cm^{-2} \cdot s^{-1} \cdot keV^{-1}}$, which is just outside the tolerance
range derived by \citet{fink94:tycho-ginga}. As in the case of model W7, $D_{norm}$ is within the estimated distance
range, but $D_{RS}$ is too high.

\paragraph{Three-dimensional deflagrations (b30\_3d\_768)}
A comparison with the spatially integrated spectrum reveals the shortcomings of three dimensional deflagrations even
more clearly than the diagnostic quantities for the line emission. In the case of model b30\_3d\_768 ($\rho_{AM}=2 \cdot
10^{-25} \, \mathrm{g \cdot cm^{-3}}$, $\beta=0.01$), the discrepancy between the predicted Fe L/Fe K$\alpha$ flux ratio
and the observed spectrum is dramatic, with the model requiring a hydrogen column density of $2.33 \cdot10^{22}\,
\mathrm{cm^{-2}}$. In fact, it can be said that the spectral adjustment is poor everywhere, and it is clear from Figure
\ref{fig-6} that no combination of $\rho_{AM}$ and $\beta$ is going to improve it. The high level of thermal continuum
in the ejecta model, which comes mostly from the large amounts of unburned C and O in the outer ejecta layers, forces a
very low value of $F_{AM}$ $(1.0\cdot10^{-2} \, \mathrm{phot\cdot cm^{-2} \cdot s^{-1} \cdot keV^{-1}})$ that is
impossible to reconcile with the \textit{Ginga} observations. As in most models that require a low value of $\rho_{AM}$,
$D_{norm}$ is too low and $D_{RS}$ is too high.

To conclude this section, we mention an unrelated, but interesting fact about the high energy continuum that we have
noticed during our attempts to approximate the integrated spectrum of the Tycho SNR. \citet{hwang02:tycho-rim} and
\citet{warren05:Tycho} found that the featureless rim emission extracted from \textit{Chandra} observations could be
fitted equally well by thermal and nonthermal models. In our integrated \textit{XMM-Newton} spectra, however, we have
found that thermal models for the continuum always underpredict the flux above 8 keV \citep[e.g., see Figure 2
in][]{bravo04:3D_review}. This is revealed by the higher effective area of \textit{XMM} at high energies, and it
constitutes yet another argument in favor of the predominantly nonthermal origin for the shocked AM emission advocated
in \citet{warren05:Tycho}.

\section{SPATIAL MORPHOLOGY OF THE LINE EMISSION} \label{sec:Spatial}

Although the focus of this paper is on the spatially integrated X-ray emission from the Tycho SNR, the spatial
distribution of this X-ray emission also contains interesting information that can shed light on the properties of the
SN ejecta and the SNR shocks. In Paper II, for instance, we proposed that partial collisionless heating at the reverse
shock could explain why the Fe K$\alpha$ emission peaks interior to Fe L and Si He$\alpha$ in both Tycho
\citep{hwang97:tycho-ASCA} and Kepler \citep{cassam03:kepler}. We note that partial collisionless electron heating has
been required by \textit{all} the models that have shown some level of success in reproducing the spatially integrated
spectrum of the Tycho SNR in the previous sections. We want to conclude our study by doing a preliminary comparison
between our models and the spatial distribution of the line emission in the Tycho SNR.

\begin{figure}

  \centering
 
  \includegraphics[scale=0.8]{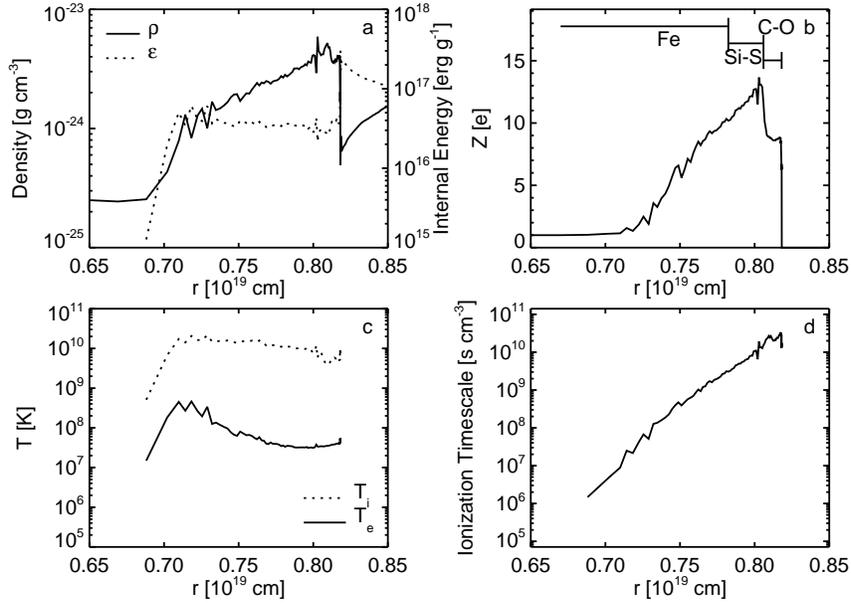}

  \caption{The radial structure of the shocked ejecta in model DDTc $(\rho_{AM}=2\cdot10^{-24}\,\mathrm{g \cdot cm^{-3}},\,\beta=0.03)$ 
    at the age of the Tycho SNR. $(a)$ Radial distribution of density $\rho$ and specific internal energy $\varepsilon$;
    $(b)$ mean number of electrons per ion, $\bar{Z}$, with an indication of the ejecta layers dominated by Fe, Si-S,
    and C-O; $(c)$ electron and ion temperatures; $(d)$ ionization timescale. The positions of the RS and CD are
    outlined by the limits of the temperature plots in $(c)$ .\label{fig-8}}

\end{figure}

\begin{figure}
  
  \centering
 
  \includegraphics[scale=0.8]{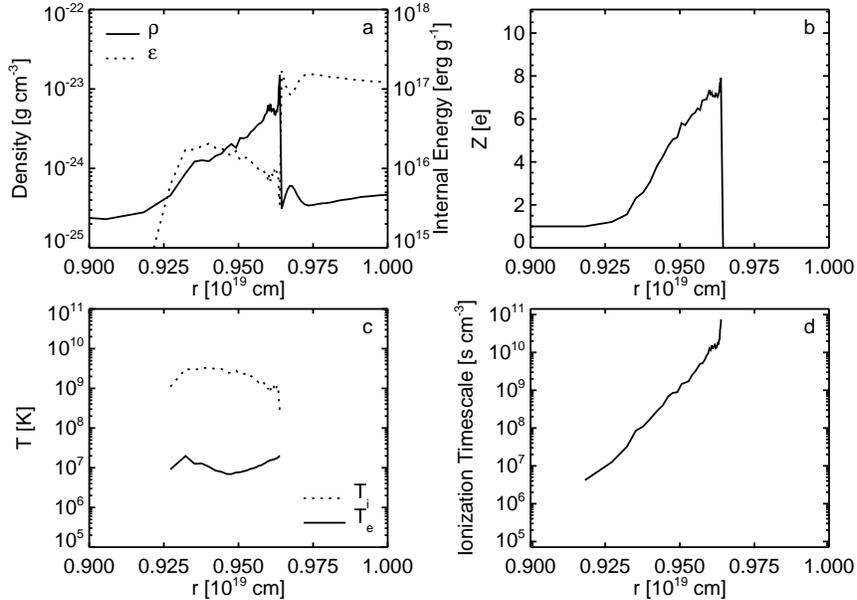}

  \caption{Same as Fig.\ref{fig-8}, but for model b30\_3d\_768 $(\rho_{AM}=2\cdot10^{-25}\,\mathrm{g \cdot cm^{-3}},\,\beta=0.01)$. 
    \label{fig-9}}

\end{figure}

We have chosen two models for this comparison: DDTc ($\rho_{AM}=2 \cdot 10^{-24} \, \mathrm{g \cdot cm^{-3}}$,
$\beta=0.03$) and b30\_3d\_768 ($\rho_{AM}=2 \cdot 10^{-25} \, \mathrm{g \cdot cm^{-3}}$, $\beta=0.01$). Both models
have a similar amount of partial collisionless electron heating at the RS, but their ejecta structures are very
different (layered vs. well-mixed), and the order of magnitude difference in $\rho_{AM}$ places them at very different
evolutionary stages. At the age of Tycho, the RS has heated $\sim1.0\,\mathrm{M_{\odot}}$ of ejecta in model DDTc, but
only $\sim0.5\,\mathrm{M_{\odot}}$ in model b30\_3d\_768. However, physical conditions in the shocked ejecta are
qualitatively similar (see Figs. \ref{fig-8} and \ref{fig-9}): $\rho$ and $n_{e}t$ increase monotonically from RS to CD,
but $T_{e}$ has local maxima at the CD (due to collisional heating in the plasma over time) and close to the RS (due to
collisionless heating at the RS).

\begin{figure}
  
  \centering
 
  \includegraphics[scale=0.6]{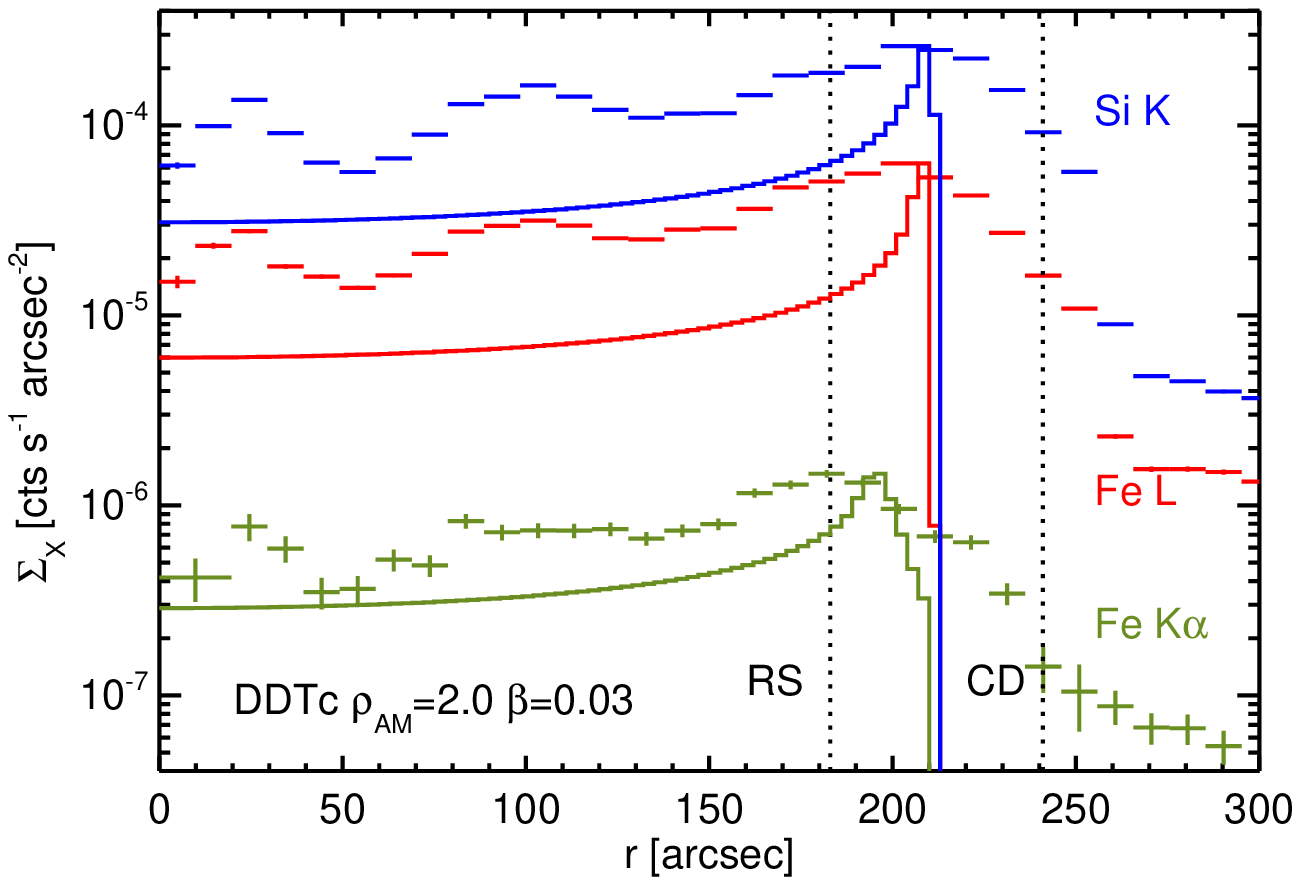}
  \includegraphics[scale=0.6]{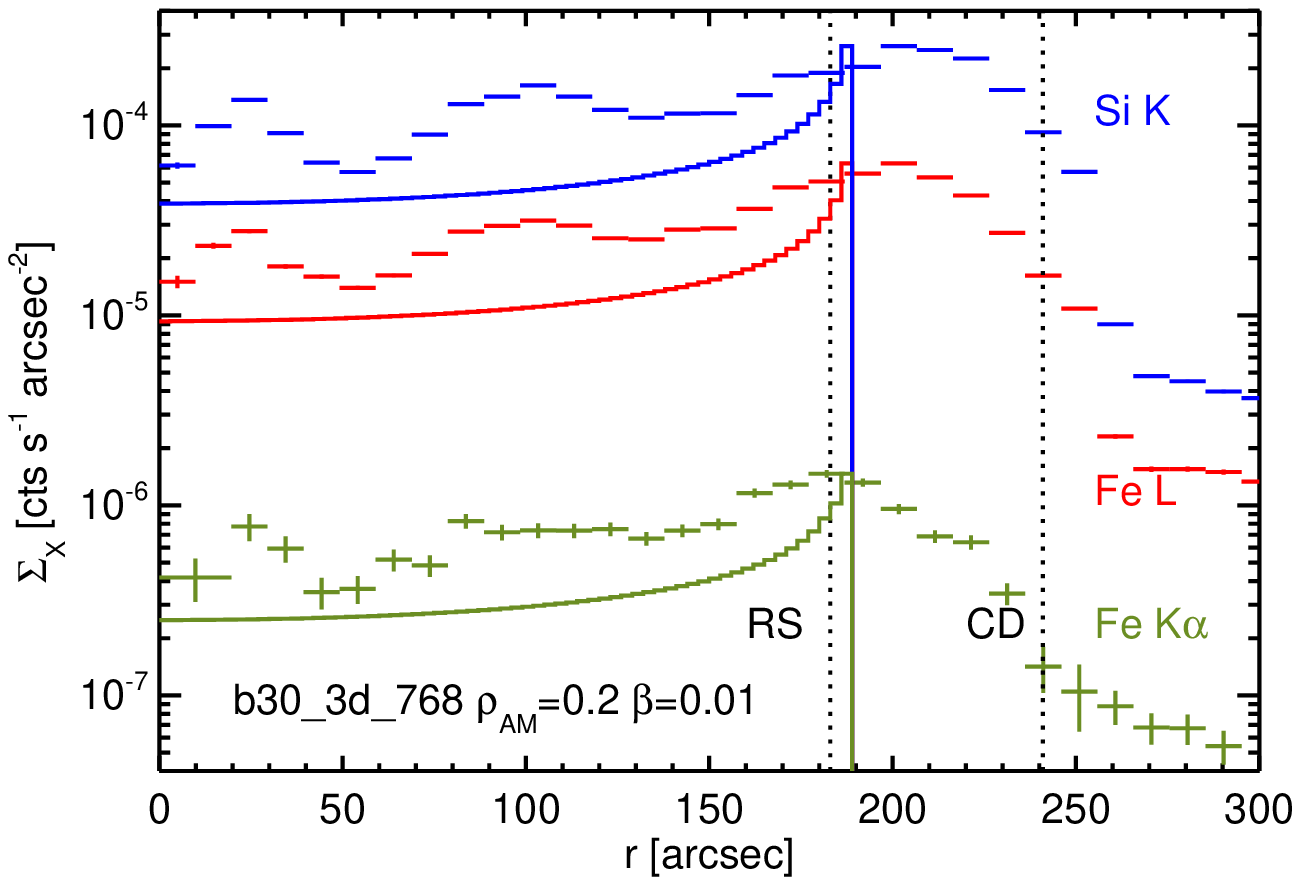}

  \caption{Radial line emissivity profiles for Fe L, Si K, and Fe K$\alpha$ in models DDTc
    (left) and b30\_3d\_768 (right), compared to the \textit{Chandra} profiles for region B. The value of $\rho_{AM}$ in
    the models is given in units of $10^{-24}\,\mathrm{g \cdot cm^{-3}}$. The locations of the RS and CD
    \citep[from][]{warren05:Tycho} have been indicated by dotted lines.\label{fig-10}}

\end{figure}

Taking advantage of the superior spatial resolution of \textit{Chandra}, we have produced radial profiles for the
emission in three spectral bands, using data from the ACIS-I CCD detectors: 0.8-0.95 keV (Fe L), 1.63-2.26 keV (Si K,
including Si He$\alpha$, Si He$\beta$, and Si Ly$\alpha$), and 6.10-6.80 keV (Fe K$\alpha$). The data are from the
$\sim150$ ks observation described in \citet{warren05:Tycho}, and were extracted from the same region B whose spatially
integrated spectrum we have used throughout the present work. For the Fe K$\alpha$ profile, the underlying continuum was
subtracted as in \citet{warren05:Tycho}. In Figure \ref{fig-10}, we plot the three line emission profiles alongside the
predictions from the two models. The model profiles have been scaled so the position of the RS coincides with the 183''
radius found by \citet{warren05:Tycho} (first vertical dotted line in Fig.  \ref{fig-10}), which is equivalent to
placing each model at the distance $D_{RS}$ listed in Table \ref{tab-6}. The model profiles have also been normalized to
the peak value in the data for each energy band. From the spectra presented in Fig.  \ref{fig-7}, it is clear that model
DDTc produces the correct spatially integrated flux in each of the three spectral bands (for $N_{H}=0.55
\cdot10^{22}\,\mathrm{cm^{-2}}$), but model b30\_3d\_768 does not.

Our one-dimensional models look very different from the data, in part because they do not extend to the observed radius
of the CD, with the radius of the outermost ejecta falling $\sim10\%$ short for model DDTc and $\sim20\%$ short for
model b30\_3d\_768. There are several processes that we have not included in our simulations and could increase the
radial extent of the ejecta, but the morphology of the CD and its power spectrum seem to indicate that Rayleigh-Taylor
instabilities (RTI) are the dominant mechanism in the Tycho SNR \citep[see \S~4, \S~7.3 and Fig. 6
in][]{warren05:Tycho}.  \citet{wang01:Ia-inst-clump} explored the development of RTI in the context of an exponential
density profile \citep[which is considered the best generic approximation to Type Ia SN ejecta,
see][]{dwarkadas98:typeIa} interacting with a uniform AM, and found an increase of $\sim10\%$ in the radius of the
outermost ejecta at the age of the Tycho SNR (see their Figure 4). This would be sufficient for model DDTc, but not for
model b30\_3d\_768. Despite the obvious limitations of our one-dimensional models, we note that model DDTc gives roughly
the right morphology: the Fe K$\alpha$ profile peaks interior to the other two, the maxima of the Fe L and Si K profiles
are at the right locations, and the Si K-shell emission extends a little bit beyond the Fe L-shell emission. This is the
result of the combination of the layered structure of the ejecta with the distribution of physical conditions in the
shocked plasma represented in Fig \ref{fig-8}. The peaks in the Fe L and Si K emission coincide with the most dense
layers with high Fe and Si abundances. The Fe K$\alpha$ emission, on the other hand, has a strong temperature
dependence, and is greatly enhanced in the hot region close to the RS produced by the collisionless electron heating at
the shock front.  For model b30\_3d\_768, on the other hand, all the predicted profiles have the same morphology.  In
this case, the peak of the Fe K$\alpha$ emission coincides with the other two because the outermost ejecta layers are
rich in Fe. A higher amount of collisionless electron heating will enhance the Fe K$\alpha$ emission close to the RS,
but this would not be beneficial for the spatially integrated spectrum of this model (see Fig. \ref{fig-6}).


\section{DISCUSSION} \label{sec:discussion}

In the present work we have compared the X-ray emission from the shocked ejecta in the Tycho SNR to the predictions from
our models based on one-dimensional hydrodynamic simulations and NEI calculations. We have focused on the spatially
integrated X-ray spectrum, but we have also taken into consideration the spatial morphology of the line emission and the
dynamics of the underlying hydrodynamic calculations. We have shown that our best models achieve a remarkable level of
success in explaining the global properties of the ejecta emission. The only inconsistency that we have found in our
comparisons is the conflict between the distance estimates $D_{norm}$ and $D_{RS}$.  In this section, we will discuss
this conflict and the impact that it has on our results, and we will summarize the case for a delayed detonation as the
physical mechanism responsible for SN 1572.

\subsection{The distance estimates: $D_{norm}$ vs. $D_{RS}$} \label{sub:limitations-models}

The distance estimates $D_{norm}$ and $D_{RS}$ never agree with each other in the models listed in Table \ref{tab-6} or,
for that matter, in any of the models that we have tested.  In the most successful class of models (the delayed
detonations), $D_{RS}$ is always within the expected 1.5-3.1 kpc range, while $D_{norm}$ is systematically larger, with
values ranging between 3.58 and 6.95 kpc. Some insight into the origin of this discrepancy can be gained by inspecting
the radial line brightness profiles given in Figure \ref{fig-10}.  The volume of X-ray emitting ejecta can be estimated
from the measured locations of the RS and CD given by \citet{warren05:Tycho} (vertical dotted lines in Fig.
\ref{fig-10}). This volume is always considerably larger than the volume occupied by the ejecta in our one-dimensional
models, assuming the RS radius in the models matches that of Tycho.  As we have discussed in the previous section, this
is probably due to the effect of Rayleigh-Taylor instabilities at the CD. Whatever the cause, spreading a fixed mass of
shocked ejecta over a larger volume will decrease the average density, reducing the intrinsic emissivity of the model
and lowering the inferred normalization distance $D_{norm}$. The value of $D_{RS}$, being determined by the hydrodynamic
evolution, should remain nearly the same \citep[see \S~3.1 in][]{wang01:Ia-inst-clump}.

The specific details of how the development of RTI will affect the emitted flux, the properties of the spatially
integrated spectrum and the spatial morphology of the line emission are complex, and a detailed study of these issues is
outside the scope of this paper. However, we want to point out here that the differences in the spatially integrated
synthetic spectrum between a one-dimensional model and a multi-dimensional model including RTI might be relatively
small. The spatially integrated emission is determined by the spectrum emitted by each individual fluid element in the SN
ejecta, which only depends on its composition, its preshock density (or equivalently, the time at which it is overrun by
the RS) and its postshock evolution. The dynamics of the RS do not change significantly in multi-D simulations including
RTI \citep[see][]{chevalier92:hydrod_instab_SNR,wang01:Ia-inst-clump}, so the density at which a fluid element with a
given composition is shocked will not be altered. After the passage of the RS, the development of RT fingers will
enhance the adiabatic expansion of the fluid element, reducing the density in the postshock evolution. The extent of
this reduction can be estimated by looking at Figure 4 in \citet{wang01:Ia-inst-clump}: at the age of the Tycho SNR, the
angle-averaged density profile affected by RTI only shows a factor 2 decrease with respect to the highest density ejecta
in a one-dimensional simulation (which is just behind the CD). In this region, the high density RT fingers penetrate
into the low density shocked AM, with the latter component filling most of the volume \citep[see Figure 2
in][]{wang01:Ia-inst-clump}, so the actual drop in density of the shocked ejecta should be less than a factor 2. We know
that the emissivity and emission measure of an X-ray plasma scale with the square of the density, while the collisional
processes in the plasma (ionization, electron heating) only have a linear density dependence (see Eqns. 1, 2, and 6 in
paper I). Therefore, it does not seem unreasonable to think that the flux emitted by the shocked ejecta may drop by
$\sim50\%$ (driving $D_{norm}$ in the DDT models into the expected range), without major changes being introduced to the
value of $n_{e}t$ and $T_{e}$ inside each fluid element and hence to the properties of the spatially integrated
spectrum. In this context, the differences in the synthetic spectra from one Type Ia SN explosion model to another
should be conserved to a large extent. However, we insist that detailed multi-dimensional simulations are required to
investigate these issues. We plan to perform such simulations in a forthcoming paper.

We conclude that the discrepancy between $D_{norm}$ and $D_{RS}$ that we have encountered might be an intrinsic
feature of one-dimensional calculations. From what we know about the structure and growth of RT fingers at the CD, their
effect on the properties of the synthetic spectra in terms of ionization timescales and electron temperatures might be
limited. From a practical point of view, the presence of RTI at the CD might modify the optimum value of $\rho_{AM}$ for a
given Type Ia SN explosion model, but it is very unlikely that it will make viable the synthetic spectrum of a model
that did not prove valid in a one-dimensional calculation.

\subsection{A delayed detonation as the origin of SN 1572}

The main result of our work is that one-dimensional delayed detonation models are successful in reproducing the
fundamental properties of the X-ray emission from the shocked ejecta in the Tycho SNR. None of the other explosion
paradigms that we have tested (prompt detonations, sub-Chandrasekhar explosions, pulsating delayed detonations,
one-dimensional deflagrations and three-dimensional deflagrations) shows a comparable level of success, and most of them
fail even simple comparisons to the observations. The delayed detonation models reproduce 12 out of the 13 diagnostic
quantities for the line emission (see Figs. \ref{fig-4}, \ref{fig-5}, and \ref{fig-6}; and Tables \ref{tab-4} and
\ref{tab-5}), and they provide good approximations to the spatially integrated \textit{XMM-Newton} spectrum at high
energies (see fig.\ref{fig-7}). All the delayed detonation models that we have tested require similar values of
$\rho_{AM}$ ($2 \cdot 10^{-24} \, \mathrm{g \cdot cm^{-3}}$ for our grid DDT models, $5 \cdot 10^{-24} \, \mathrm{g
  \cdot cm^{-3}}$ for the `off-grid' model 5p0z22.25) and $\beta$ (between 0.01 and 0.1). These values lead to estimates
for the distance $D_{RS}$, the nonthermal flux from the shocked AM $F_{AM}$, and (in the case of model DDTc) the
hydrogen column density $N_{H}$, which are fully consistent with the available observational constraints. We emphasize
that, by taking these estimates into account, we are evaluating our models from a global point of view, not based on the
properties of their synthetic spectra alone. We have also shown that the spatial morphology of the line emission
predicted by delayed detonation models with partial collisionless electron heating is roughly consistent with the
\textit{Chandra} observations of the Tycho SNR, at least within the limitations of one-dimensional models. The
inconsistency between the distance estimates from matching the RS radius ($D_{RS}$) and from the normalization of the
spectrum ($D_{norm}$) that we found in all the models could also be related to the limitations of our one-dimensional
hydrodynamic calculations. A detailed study of the effect of the Rayleigh-Taylor instability on the X-ray emission from
the shocked ejecta would be necessary to gain further insight on the spatial morphology of the line emission and the
values of $D_{norm}$.

We believe that the unavoidable limitations of one-dimensional simulations do not compromise our results concerning
delayed detonation models (see discussion in \ref{sub:limitations-models}). This is specially true because we have found
no alternative paradigm with an acceptable level of success whose predictions could be reconciled with the observations
by invoking the limited modifications that we expect from multi-dimensional effects like RTI. We note that our
conclusions do not agree with those of \citet{sorokina04:typeIasnrs}, who show a preference for three-dimensional
deflagrations for the Tycho SNR. However, these authors only consider two deflagration models (W7 and a model similar to
b30\_3d\_768), and their results are based on a very simple qualitative comparison between models and observations. To
evaluate the merits of our slightly different approaches on an equal footing, synthetic spectra from delayed detonation
models should be produced using the methods of \citeauthor{sorokina04:typeIasnrs}, and they should be compared to the
observations in a rigorous, quantitative way.

\subsection{The case for model DDTc ($\rho_{AM}=2 \cdot 10^{-24} \,\mathrm{g \cdot cm^{-3}}$, $\beta=0.03$)} \label{sub:case-model-ddtc}

Among the delayed detonation models that we have tested, model DDTc ($\rho_{AM}=2 \cdot 10^{-24} \, \mathrm{g \cdot
  cm^{-3}}$, $\beta=0.03$) gives by far the best approximation to the spatially integrated X-ray spectrum. The
performance of this model is surprisingly good: despite the limitations of our one-dimensional calculations and the
large uncertainties in the atomic data used to generate the synthetic spectra, the model can reproduce the fluxes and
centroids of almost all the lines from O, Si, S, Ar, Ca and Fe in the shocked ejecta between 0.6 and 7.0 keV (see Fig.
\ref{fig-7}). Our preference for this model is based on the constraints posed by the spectrum at low energies (below 1
keV) on the main parameter involved in delayed detonation explosions, the transition density from deflagration to
detonation $\rho_{tr}$. Model DDTc matches both the Fe L/Fe K$\alpha$ flux ratio and the O Ly$\alpha$ emission for a
reasonable value of $N_{H}$ ($0.55\cdot10^{22}\,\mathrm{cm^{-2}}$), unlike models with higher or lower $\rho_{tr}$ such
as DDTa or DDTe. The second most successful delayed detonation model, 5p0z22.25, although calculated using a different
SN explosion code, is very similar to DDTc. Both models have similar values of $\rho_{tr}$ ($2.2 \cdot 10^{7}\,\mathrm{g
  \cdot cm^{-3}}$ for DDTc and $2.5 \cdot 10^{7}\,\mathrm{g \cdot cm^{-3}}$ for 5p0z22.25), similar kinetic energies
($1.16$ and $1.20 \cdot 10^{51}$ erg), and they synthesize similar amounts of most elements (Fe: 0.80 and 0.74
$\mathrm{M_{\odot}}$, Si: 0.17 and 0.22 $\mathrm{M_{\odot}}$, S: 0.13 and 0.12 $\mathrm{M_{\odot}}$).

The fact that the model agrees so well with the observed spectrum implies that the bulk properties of the synthetic
spectrum (i.e., emission measures and emission measure averaged temperatures and ionization timescales) are roughly
correct for the most prominent elements, and can be considered representative of the state of the shocked ejecta in the
Tycho SNR. The spectral adjustment provided by the model, however, is not perfect. As we discussed in
\S~\ref{sub:comparing}, the centroid of the Ca He$\alpha$ blend is underpredicted by the model. A close examination of
the residuals in Fig. \ref{fig-7} reveals that the centroids of the Si He$\alpha$, S He$\alpha$, Ar He$\alpha$ and Fe
K$\alpha$ blends in the synthetic spectrum are also offset towards lower energies with respect to the data. The shape of
these residuals is suggestive of a gain shift, but we have verified that this is not the case - the shift is indeed
real, and indicates that the plasma in the model is slightly underionized. Increasing the value of $\rho_{AM}$ gradually
between $2.0$ and $5.0 \cdot 10^{-24} \, \mathrm{g \cdot cm^{-3}}$ might solve this issue, but given the limitations of
one-dimensional models discussed in \S~\ref{sub:limitations-models}, we do not believe that such refinements are
justified.

The mean velocity of the shocked ejecta in this model is $\sim2\cdot10^{8}\,\mathrm{cm \cdot s^{-1}}$. For a radius of
$\sim7\cdot10^{18}\,\mathrm{cm}$ (see Fig. \ref{fig-8}), this gives an expansion of $\sim0.09\%\,\mathrm{yr^{-1}}$. This
is in very good agreement with the expansion parameters of the interior features of the SNR measured in radio
\citep[$\eta\simeq0.44$,][]{reynoso99:Tycho_VLA_II} and X-rays
\citep[$\eta\simeq0.45$,][]{hughes00:Expansion_Tycho_X_Rays}, which translate into a percentual expansion of of
$\sim0.11\%\,\mathrm{yr^{-1}}$. We note that, when the peaks of the Si He$\alpha$ blend are aligned, the width of the
blend seems larger in the data than in the synthetic spectrum, suggesting a Doppler broadening of the line emission in
the shocked ejecta. This would certainly have interesting implications for the dynamics of the Tycho SNR, but a
quantitative analysis of this effect would be challenging given the spectral resolution of the CCD cameras onboard
\textit{Chandra} and \textit{XMM-Newton}.

We conclude our comments with a reference to the historical observations of SN 1572. The light curve calculated for
model DDTc yields a bolometric magnitude at light curve maximum $\mathcal{M}_{max}$ of -19.51 mag (see Table 3 in Paper
II), which is at the upper limit of the $\mathcal{M}_{V}=-19.24-5log(D/3.0 \mathrm{kpc})\pm0.42$ range derived by
\citet{ruiz-lapuente04:TychoSN} taking the value of 2.8 kpc for $D$ proposed by this author, but $0.17$ mag too bright
at $D_{RS}=2.59$ kpc. Given the fact that the estimates of \citeauthor{ruiz-lapuente04:TychoSN} are based on naked eye
observations, and the large range of peak bolometric magnitudes in our Type Ia SN model grid (more than 2 mag, see Table
1 in Paper I), this can be considered a good agreement. An explosion that synthesizes $\sim0,8\,\mathrm{M_{\odot}}$ of
$\mathrm{^{56}Ni}$ (like model DDTc) is consistent with the hypothesis that SN 1572 was a `normal' Type Ia SN, and not a
subluminous event.

\section{CONCLUSIONS} \label{sec:Conclusions}

In the present work, we have compared synthetic X-ray spectra for the ejecta emission in Type Ia SNRs with the X-ray
observations of the Tycho SNR, with the goal of constraining the explosion mechanism that gave birth to the supernova of
1572. In our comparisons, we have found a very strong bias towards delayed detonation models, to the point that no other
Type Ia SN explosion paradigm has even come close to reproducing the fundamental properties of the X-ray emission from
the ejecta in the Tycho SNR. This result has a high significance, and we believe that it is not affected by the
limitations of our one-dimensional models.

Among the delayed detonation models that we have tested, we obtained the best results for model DDTc, expanding into a
uniform AM of $\rho_{AM}=2 \cdot 10^{-24} \, \mathrm{g \cdot cm^{-3}}$, and with a modest amount of collisionless
heating at the RS ($\beta=0.03$). At the age of the Tycho SNR, the synthetic spectrum calculated for this model can
account for almost all the fundamental properties of the ejecta emission in the \textit{XMM-Newton} observation,
including the line emission from O, Si, S, Ar, Ca, and Fe between 0.6 and 7.0 keV. In agreement with the results of
\citet{warren05:Tycho}, the spectrum only requires a nonthermal contribution from the AM, with a flux of $8.1 \cdot
10^{-2}\,\mathrm{phot \cdot cm^{-2} \cdot s^{-1} \cdot keV^{-1}}$, which is within the range determined by
\citet{fink94:tycho-ginga}, and a hydrogen column density of $0.55\cdot10^{22}\,\mathrm{cm^{-2}}$, which is also
compatible with previous observations \citep[e.g.][]{hwang98:tycho-Feemission}. By matching the radius of the RS in the
model to the 183'' radius found by \citet{warren05:Tycho}, we obtain a distance estimate of 2.59 kpc, within the range
determined by \citet{smith91:six_balmer_snrs}. The distance estimated from the normalization of the emitted spectrum is
not consistent with this, but the origin of this discrepancy is probably in the inherent limitations of the
one-dimensional hydrodynamic calculations that underlie our synthetic spectra. The spatial morphology of the line
emission is qualitatively consistent with the \textit{Chandra} observations of the Tycho SNR, although multi-D
simulations are required for a quantitative comparison with observations. We conclude that model DDTc represents the
closest approximation to the structure of the ejecta in the Tycho SNR that we can find within our grid of synthetic
spectra. The deflagration-to-detonation transition density of this model is $\rho_{tr} = 2.2 \cdot 10^{7}\,\mathrm{g
  \cdot cm^{-3}}$, its kinetic energy is $E_{k}=1.16 \cdot 10^{51}$ erg, and it synthesizes $0.12\,\mathrm{M_{\odot}}$
of C and O, $0.17\,\mathrm{M_{\odot}}$ of Si, $0.13\,\mathrm{M_{\odot}}$ of S, $0.033\,\mathrm{M_{\odot}}$ of Ar,
$0.038\,\mathrm{M_{\odot}}$ of Ca, and $0.80\,\mathrm{M_{\odot}}$ of Fe. We note that a future analysis based on more
sophisticated (e.g., multi-D) SNR models and/or a different grid of Type Ia explosion models might lead to a revision
of the specific values for the kinetic energy and nucleosynthetic yields that we cite here, but we do not anticipate
major changes.

Our results have interesting implications for the analysis of X-ray observations of Type Ia SNRs. Our models allow to
consider the emission from all the elements in the ejecta at once, and they are characterized by very few parameters
with a clear physical meaning: the age of the SNR $t$, the AM density $\rho_{AM}$, the efficiency of collisionless
electron heating at the RS $\beta$, and the explosion mechanism. These models constitute a promising way to approach the
analysis of the ejecta emission, foregoing many of the ambiguities inherent to more conventional techniques based on
spectral fitting (for a discussion, see Rakowski et al., ApJ, submitted). Futhermore, by interpreting the X-ray spectrum
of the SNR in terms of realistic supernova explosion models, it is possible to distinguish between remnants originated
by core collapse or thermonuclear supernovae with a higher level of confidence. In future papers, we plan to apply our
grid of synthetic spectra to the other prototype Galactic Type Ia SNR, SN 1006, and to improve our models by including
multi-dimensional effects like Rayleigh-Taylor instabilities.

The implications for the study of SN explosions are also interesting. The X-ray observations of SNRs open a new window
onto the structure of the SN ejecta, totally independent of the optical, ultraviolet and infrared light curves and
spectra of SNe. The two approaches are complementary, because they involve completely different challenges and
constraints: modeling the radiative transfer in the expanding ejecta in the case of the spectral evolution of Type Ia
SNe \citep[see][and references therein]{branch05:SN1994D} vs. modeling the dynamics of the shocked plasma in Type Ia
SNRs.  One advantage of SNRs is that the emitting plasma is optically thin, and all the elements can be seen in the
X-rays at the same time. This makes the success of delayed detonation models for the Tycho SNR all the more significant,
and the agreement with previous studies of Type Ia SNe, which also showed a preference for delayed detonations
\citep[e.g.][]{hoeflich96:nosubCh}, more compelling. It is worth noting that we have only examined the properties of the
ejecta emission from the Tycho SNR on the western side of the remnant, where they are reasonably
homogeneous. Significant spatial inhomogeneities exist in the eastern side \citep{vancura95:Tycho}, which could provide
important clues about asymmetries in Type Ia SN explosions.

We conclude with a reminder that, despite their apparent success, delayed detonations are just a phenomenological model
for Type Ia SNe. Since this explosion paradigm was first proposed by \citet{khokhlov91:ddt} \footnote{A pulsating regime
  of detonation was proposed in 1974 by \citeauthor{ivanova74:pulsating_detonation}, but see the discussion in p. 230 of
  \citet{khokhlov93:lightcurves}}, all attempts to provide a physically consistent mechanism that can explain the
transition from deflagration to detonation have failed. Some simulations have explored delayed detonation explosions in
three dimensions \citep{gamezo05:DDT3D}, but without providing any further insight into the physical mechanism that
governs the transition. Yet, our results provide unambiguous evidence that the ejecta of the Type Ia SN that originated
the Tycho SNR are stratified, which implies that a detonation was involved at some stage in the physics of the
explosion. Very recently, two new explosion mechanisms for Type Ia SNe have been proposed that lead to detonations in a
self-consistent way: gravitationally confined detonations \citep{plewa04:GCD} and pulsating reverse detonations (Bravo
\& Garc\'ia-Senz, ApJ, submitted). Both these paradigms result in a stratified structure for the SN ejecta, which is
qualitatively compatible with our results, but a quantitative comparison is required to evaluate their merits in the
context of the X-ray emission from Type Ia SNRs. There is clearly much work left to be done before the physics of Type
Ia SN explosions can be completely understood, and we expect the X-ray observations of young, ejecta-dominated Type Ia
SNRs to make a significant contribution to this crucial issue.

\acknowledgements This work is dedicated to the memory of Vicente Badenes, the grandfather of Carles, who passed away
on March 9, 2005.

We are grateful to Ken'ichi Nomoto, Peter H\"{o}flich, Claudia Travaglio and Wolfgang Hillebrandt for providing their
Type Ia explosion models W7, 5p0z22.25 and b30\_3d\_768, and to Manuel Bautista for providing corrected tables for Fe
K$\alpha$ inner-shell electron collisional excitation cross sections. We wish to thank Martin Laming and Cara Rakowski
for many stimulating discussions on the SN-SNR connection, and Craig Gordon, Bryan Irby, and the rest of the XSPEC
helpdesk for assistance with running local models in XSPEC 12. Part of this work was conducted during a stay of C.B. at
NASA-GSFC in the summer of 2003, funded by the Generalitat de Catalunya.  C.B. would also like to acknowledge support
from Generalitat de Catalunya (grant 2000FI 00376) and IEEC in Barcelona, and from grant GO3-4066X from SAO at Rutgers.
K.J.B. is supported by NASA through grant NAG 5-7153. E.B. has received support from the DURSI of the Generalitat de
Catalunya and the Spanish DGICYT grants AYA2002-0494-C03 and AYA2004-06290-C02-02.
  


\clearpage
\centering
\begin{deluxetable}{clccccc}
  \rotate
  \tablewidth{0pt}
  \tabletypesize{\scriptsize}
  \tablecaption{Line fluxes and centroids in the Tycho SNR\label{tab-1}}
  \tablecolumns{7}
  \tablehead{
    \colhead{} &
    \colhead{} &
    \colhead{} &
    \multicolumn{2}{c}{Region A} &
    \multicolumn{2}{c}{Region B} \\
    \colhead{} &
    \colhead{} &
    \colhead{Expected} &
    \colhead{Centroid} &
    \colhead{Flux} &
    \colhead{Centroid} &
    \colhead{Flux} \\
    \colhead{Line} &
    \colhead{Ion and Transition} &
    \colhead{Centroid (keV)} &
    \colhead{(keV)} &
    \colhead{$(10^{-3}\mathrm{phot\cdot cm^{-2}\cdot s^{-1}})$} &
    \colhead{(keV)} &
    \colhead{$(10^{-3}\mathrm{phot\cdot cm^{-2}\cdot s^{-1}})$}
  }
  \startdata
  Si He$\alpha$ & $\mathrm{Si^{+12}},\, n=2\rightarrow n=1$ & $\sim1.86$ &
  $1.8584_{-0.0002}^{+0.0003}$ \tablenotemark{a} & $30.6_{-0.1}^{+0.2}$ & $1.8578_{-0.0002}^{+0.0004}$ & $18.0 \pm 0.1$ \\
  Si He$\beta$ & $\mathrm{Si^{+12}},\,1s3p\rightarrow1s^{2}$ & $2.182$ &
  $2.185_{-0.001}^{+0.002}$ & $2.60_{-0.04}^{+0.05}$ & $2.184_{-0.001}^{+0.002}$ & $1.54 \pm 0.04$ \\
  Si He$\gamma$ & $\mathrm{Si^{+12}},\,1s4p\rightarrow1s^{2}$ & $2.294$ &
  ... \tablenotemark{b} & $0.35\times\mathrm{Si\, He\beta}$ & ... & $0.35\times\mathrm{Si\, He\beta}$ \\
  Si Ly$\alpha$ & $\mathrm{Si^{+13}},\,2p\rightarrow1s$ & $2.006$ &
  ... & $<1.04$ \tablenotemark{c}& ... & $<0.58$ \tablenotemark{c}\\
  Si Ly$\beta$ & $\mathrm{Si^{+13}},\,3p\rightarrow1s$ & $2.377$ &
  ... & $0.14\times\mathrm{Si\, Ly\alpha}$ & ... & $0.14\times\mathrm{Si\, Ly\alpha}$ \\
  S He$\alpha$ & $\mathrm{S^{+14}},\, n=2\rightarrow n=1$ & $\sim2.45$ &
  $2.447 \pm 0.001$ & $8.14_{-0.11}^{+0.08}$ & $2.447 \pm 0.001$ & $4.75_{-0.08}^{+0.06}$ \\
  S He$\beta$ & $\mathrm{S^{+14}},\,1s3p\rightarrow1s^{2}$ & $2.884$ &
  ... & $0.52 \pm 0.03$ & ... & $0.29 \pm 0.02$ \\
  S He$\gamma$ & $\mathrm{S^{+14}},\,1s4p\rightarrow1s^{2}$ & $3.033$ &
  ... & $0.40\times\mathrm{S\, He\beta}$ & ... & $0.40\times\mathrm{S\, He\beta}$ \\ 
  S Ly$\alpha$ & $\mathrm{S^{+15}},\,2p\rightarrow1s$ & $2.623$ &
  ... & $<0.05$ \tablenotemark{c}& ... & $<0.02$ \tablenotemark{c}\\ 
  Ar He$\alpha$ & $\mathrm{Ar^{+16}},\, n=2\rightarrow n=1$ & $\sim3.1$ &
  $3.131_{-0.004}^{+0.003}$ & $0.68_{-0.03}^{+0.04}$ & $3.124_{-0.005}^{+0.003}$ & $0.38_{-0.02}^{+0.03}$ \\
  Ar He$\beta$ & $\mathrm{Ar^{+16}},\,1s3p\rightarrow1s^{2}$ & $3.685$ &
  ... & $0.06 \pm 0.02$ & ... & $0.02_{-0.01}^{+0.02}$ \\ 
  Ar He$\gamma$ & $\mathrm{Ar^{+16}},\,1s4p\rightarrow1s^{2}$ & $3.875$ &
  ... & $0.57\times\mathrm{Ar\, He\beta}$ & ... & $0.57\times\mathrm{Ar\, He\beta}$ \\
  Ca He$\alpha$ & $\mathrm{Ca^{+18}},\, n=2\rightarrow n=1$ & $\sim3.88$ &
  $3.879 \pm 0.009$ & $0.24 \pm 0.03$ & $3.867_{-0.015}^{+0.013}$ & $0.15 \pm 0.02$ \\
  Fe K$\alpha$ & Several, $n=2\rightarrow n=1$ & $\sim6.45$ &
  $6.462 \pm 0.009$ & $0.26 \pm 0.02$ & $6.456 \pm 0.013$ & $0.15 \pm 0.02$
  \enddata
  
  \tablenotetext{a}{The limits given are the formal $90\%$ confidence ranges $(\Delta \chi^{2}=2.706)$.}
  \tablenotetext{b}{Centroids marked as ... were not fitted.}
  \tablenotetext{c}{For the Si and S Ly$\alpha$ lines, only the 3$\sigma$ upper limits to the flux are given.}


\end{deluxetable}

\clearpage
\centering
\begin{deluxetable}{ccc}
  \tablewidth{0pt}
  \tabletypesize{\scriptsize}
  \tablecaption{Diagnostic line ratios for the Tycho SNR\label{tab-2}}
  \tablecolumns{3}
  \tablehead{
    \colhead{Line Ratio} & 
    \multicolumn{2}{c}{Fitted Values} \\
    \colhead{} & 
    \colhead{Region A} & 
    \colhead{Region B}
  }
  \startdata
  SiHe($\beta+\gamma$)/SiHe$\alpha$ & $0.115\pm0.003$ \tablenotemark{a}& $0.116\pm0.004$ \\
  SiLy$\alpha$/SiHe$\alpha$ &  $<0.034$ & $<0.032$ \\
  SHe($\beta+\gamma$)/SHe$\alpha$ & $0.090\pm0.006$ & $0.084\pm0.008$ \\ 
  SLy$\alpha$/SHe$\alpha$ & $<0.006$ & $<0.004$ \\
  SiHe$\alpha$/SHe$\alpha$ & $3.76\pm0.07$ & $3.80\pm0.09$ \\
  ArHe$\alpha$/SHe$\alpha$ & $0.084\pm0.006$ & $0.080\pm0.007$ \\
  CaHe$\alpha$/SHe$\alpha$ & $0.029\pm0.004$ & $0.032\pm0.006$ \\
  FeK$\alpha$/SHe$\alpha$ & $0.032\pm0.003$ & $0.032\pm0.004$ 
  \enddata
  
  \tablenotetext{a}{For line fluxes with asymmetric $90\%$ confidence ranges, the error in the line flux ratios has been
    calculated assuming symmetric confidence ranges with the largest of the two deviations.}

\end{deluxetable}




\centering
\begin{deluxetable}{cc}
  \tabletypesize{\scriptsize}
  \tablewidth{0pt}
  \tablecaption{Extraction energy windows for the model spectra \label{tab-3}}
  \tablecolumns{2}
  \tablehead{
    \colhead{Line or blend} & \colhead{Energy Window (keV)}
  }
  \startdata
  Si He$\alpha$ & $1.80-1.90$ \\
  Si He$\beta$ & $2.17-2.20$ \\
  Si He$\gamma$ & $2.28-2.32$ \\
  Si Ly$\alpha$ & $1.97-2.02$ \\ 
  S He$\alpha$ & $2.35-2.50$ \\
  S He$\beta$ & $2.87-2.90$ \\ 
  S He$\gamma$ & $3.03-3.05$ \\ 
  S Ly$\alpha$ & $2.61-2.64$ \\
  Ar He$\alpha$ & $3.05-3.16$ \\
  Ar He$\beta$\tablenotemark{a} & $3.67-3.70$ \\
  Ca He$\alpha$\tablenotemark{b} & $3.70-3.93$ \\ 
  Fe K$\alpha$\tablenotemark{c} & $6.20-6.90$ 
  \enddata

  \tablenotetext{a}{Might include some Ca K$\alpha$ lines around 3.7 keV.}
  \tablenotetext{b}{Might include some Ca K$\alpha$ lines around 3.7 keV and 
    a contribution from Ar He$\gamma$ at 3.88 keV.}
  \tablenotetext{c}{Might include some Fe He$\alpha$ lines around 6.67 keV.}

\end{deluxetable}

\clearpage 

\centering

\begin{deluxetable}{lcccccccccc}
  \rotate
  \tablewidth{0pt}
  \tabletypesize{\scriptsize}
  \tablecaption{Models vs. observations: line flux ratios \label{tab-4}}
  \tablecolumns{11}
  \tablehead{
    \colhead{} &
    \multicolumn{2}{c}{Parameters} &
    \multicolumn{8}{c}{Line Flux Ratios} \\
    \colhead{} &
    \colhead{$\rho_{AM}$} &
    \colhead{$\beta$} &
    \colhead{Si He($\beta+\gamma$)/} &
    \colhead{Si Ly$\alpha$/} &
    \colhead{S He($\beta+\gamma$)/} &
    \colhead{S Ly$\alpha$/} &
    \colhead{Si He$\alpha$/} &
    \colhead{Ar He$\alpha$/} &
    \colhead{Ca He$\alpha$/} &
    \colhead{Fe K$\alpha$/}  \\
    \colhead{Model} & 
    \colhead{($10^{-24}\,\mathrm{g \cdot cm^{-3}}$)} & 
    \colhead{} & 
    \colhead{Si He$\alpha$} &
    \colhead{Si He$\alpha$} &
    \colhead{S He$\alpha$} &
    \colhead{S He$\alpha$} &
    \colhead{S He$\alpha$} &
    \colhead{S He$\alpha$} &
    \colhead{S He$\alpha$} &
    \colhead{S He$\alpha$} 
  }
  \startdata
  DDTa & 2.0 & 0.01 &
  $\checkmark$ & $\checkmark$ & $\checkmark$ & $\checkmark$ & 
  $\checkmark$ & $\checkmark$ & $\checkmark$ & $\checkmark$ \\
  DDTc & 2.0 & 0.01 $\div$ 0.1 &
  $\checkmark$ & $\checkmark$ & $\checkmark$ & $\checkmark$ & 
  $\checkmark$ & $\checkmark$ & $\checkmark$ & $\checkmark$ \\
  DDTe & 2.0 & 0.1 &
  $\checkmark$ & $\checkmark$ & $\checkmark$ & $\checkmark$ & 
  $\checkmark$ & $\checkmark$ & $\checkmark$ & $\checkmark$ \\
  5p0z22.25 & 5.0 & 0.01 $\div$ 0.1 &
  $\checkmark$ & $\checkmark$ & $\checkmark$ & $\checkmark$ & 
  $\checkmark$ & $\checkmark$ & $\checkmark$ & $\checkmark$ \\
  \hline
  PDDa & 0.2 & $\beta_{min}$ $\div$ 0.01 &
  $\checkmark$ & $\checkmark$ & $\checkmark$ & $\checkmark$ & 
  $\checkmark$ & $\checkmark$ & $\checkmark$ & $\checkmark$ \\
  PDDc & 0.2 & 0.01 &
  $\checkmark$ & $\checkmark$ & $\checkmark$ & $\checkmark$ & 
  $\checkmark$ & $\checkmark$ & $\checkmark$ & $\sim$ \\
  PDDe & 0.2 & 0.1 &
  $\checkmark$ & $\checkmark$ & $\checkmark$ & $\dag$ & 
  $\checkmark$ & $\checkmark$ & $\sim$ & $\checkmark$ \\
  \hline
  DEFa & 5.0 & 0.1 &
  $\checkmark$ & $\dag$ & $\checkmark$ & $\dag$ &  
  $\checkmark$ & $\checkmark$ & $\nexists$ & $\checkmark$ \\
  DEFc & 5.0 & 0.01 &
  $\checkmark$ & $\dag$ & $\checkmark$ & $\dag$ & 
  $\sim$ & $\checkmark$ & $\checkmark$ & $\checkmark$ \\
  DEFf & 2.0 & 0.01 $\div$ 0.1 &
  $\checkmark$ & $\dag$ & $\checkmark$ & $\dag$ & 
  $\checkmark$ & $\checkmark$ & $\nexists$ & $\checkmark$ \\
  \hline
  W7 & 0.5 & 0.1 &
  $\checkmark$ & $\checkmark$ & $\checkmark$ & $\checkmark$ & 
  $\checkmark$ & $\sim$ & $\nexists$ & $\checkmark$ \\
  SCH & 0.5 & 0.01 &
  $\checkmark$ & $\checkmark$ & $\checkmark$ & $\checkmark$ & 
  $\checkmark$ & $\dag$ & $\nexists$ & $\sim$ \\
  DET & 2.0 & $\beta_{min}$ &
  $\nexists$ & $\nexists$ & $\nexists$ & $\nexists$ & 
  $\nexists$ \tablenotemark{a} & $\nexists$ & $\nexists$ & $\nexists$ \tablenotemark{a} \\
  \hline
  B30U & 2.0 & 0.1 &
  $\checkmark$ & $\checkmark$ & $\checkmark$ & $\checkmark$ & 
  $\dag$ & $\nexists$ & $\nexists$ & $\dag$ \\
  b30\_3d\_768 & 0.2 & 0.01 &
  $\checkmark$ & $\checkmark$ & $\checkmark$ & $\checkmark$ & 
  $\checkmark$ & $\nexists$ & $\nexists$ & $\checkmark$ 
  \enddata  
  
  \tablecomments{$\checkmark$: within tolerance range; $\sim$: marginal (on the border of the tolerance range);
    $\dag$: discrepant; $\nexists$: line is too weak to calculate a flux or is absent from the synthetic spectrum.}
  
  \tablenotetext{a}{The Si He$\alpha$ and Fe K$\alpha$ blends are both present in the synthetic spectrum, but the
    S He$\alpha$ blend is too weak to calculate these line flux ratios.}

\end{deluxetable}

\clearpage

\centering

\begin{deluxetable}{lcccccccc}
  \rotate
  \tablewidth{0pt}
  \tabletypesize{\scriptsize}
  \tablecaption{Models vs. observations: line centroids \label{tab-5}}
  \tablecolumns{9}
  \tablehead{
    \colhead{} &
    \multicolumn{2}{c}{Parameters} &
    \multicolumn{5}{c}{Line Centroids} &
    \colhead{Observables} \\
    \colhead{} &
    \colhead{$\rho_{AM}$} &
    \colhead{$\beta$} &
    \multicolumn{5}{c}{} &
    \colhead{Reproduced} \\
    \colhead{Model} & 
    \colhead{($10^{-24}\,\mathrm{g \cdot cm^{-3}}$)} & 
    \colhead{} & 
    \colhead{Si He$\alpha$} &
    \colhead{S He$\alpha$} &
    \colhead{Ar He$\alpha$} &
    \colhead{Ca He$\alpha$} &
    \colhead{Fe K$\alpha$} &
    \colhead{(out of 13)}
  }
  \startdata
  DDTa & 2.0 & 0.01 & 
  $\checkmark$ & $\checkmark$ & $\checkmark$ & $\sim$ & $\checkmark$ & 12 \\
  DDTc & 2.0 & 0.01 $\div$ 0.1 & 
  $\checkmark$ & $\checkmark$ & $\checkmark$ & $\dag$ & $\checkmark$ & 12 \\
  DDTe & 2.0 & 0.1 & 
  $\checkmark$ & $\checkmark$ & $\checkmark$ & $\dag$ & $\dag$ & 11 \\
  5p0z22.25 & 5.0 & 0.01 $\div$ 0.1 & 
  $\checkmark$ & $\checkmark$ & $\checkmark$ & $\sim$ & $\checkmark$ & 12 \\
  \hline
  PDDa & 0.2 & $\beta_{min}$ $\div$ 0.01 & 
  $\checkmark$ & $\checkmark$ & $\checkmark$ & $\sim$ & $\checkmark$ & 12 \\
  PDDc & 0.2 & 0.01 & 
  $\checkmark$ & $\checkmark$ & $\checkmark$ & $\dag$ & $\dag$ & 11 \\
  PDDe & 0.2 & 0.1 & 
  $\checkmark$ & $\checkmark$ & $\checkmark$ & $\dag$ & $\dag$ & 9 \\
  \hline
  DEFa & 5.0 & 0.1 &
  $\checkmark$ & $\checkmark$ & $\checkmark$ & $\nexists$ & $\checkmark$ & 9 \\
  DEFc & 5.0 & 0.01 & 
  $\checkmark$ & $\checkmark$ & $\checkmark$ & $\checkmark$ & $\dag$ & 9 \\
  DEFf & 2.0 & 0.01 $\div$ 0.1 & 
  $\checkmark$ & $\checkmark$ & $\checkmark$ & $\nexists$ & $\checkmark$ & 9 \\
  \hline
  W7 & 0.5 & 0.1 & 
  $\checkmark$ & $\checkmark$ & $\checkmark$ & $\nexists$ & $\dag$ & 9 \\
  SCH & 0.5 & 0.01 & 
  $\checkmark$ & $\checkmark$ & $\checkmark$ & $\nexists$ & $\sim$ & 8 \\
  DET & 2.0 & $\beta_{min}$ & 
  $\checkmark$ & $\nexists$ & $\nexists$ & $\nexists$ & $\checkmark$ & 2 \\
  \hline
  B30U & 2.0 & 0.1 & 
  $\checkmark$ & $\checkmark$ & $\nexists$ & $\nexists$ & $\dag$ & 6 \\
  b30\_3d\_768 & 0.2 & 0.01 & 
  $\checkmark$ & $\checkmark$ & $\nexists$ & $\nexists$ & $\dag$ & 8
  \enddata  

  \tablecomments{$\checkmark$: within tolerance range; $\sim$: marginal (on the border of the tolerance range);
    $\dag$: discrepant; $\nexists$: line is too weak to calculate a centroid or is absent from the synthetic spectrum.}

\end{deluxetable}

\clearpage

\centering

\begin{deluxetable}{ccccccccc}
  \tablewidth{0pt}
  \tabletypesize{\scriptsize}
  \tablecaption{Type Ia explosion models for the Tycho SNR \label{tab-6}}
  \tablecolumns{9}
  \tablehead{
    \colhead{} &
    \colhead{} &
    \colhead{} &
    \colhead{} &
    \colhead{$norm_{AM}$ \tablenotemark{a}} &
    \colhead{$norm_{ej}$ \tablenotemark{a}} &
    \colhead{$F_{AM}$} &
    \colhead{} &
    \colhead{} \\
    \colhead{Model} &
    \colhead{$\rho_{AM}$} &
    \colhead{$\beta$} &
    \colhead{$N_{H}$ \tablenotemark{a}} &
    \colhead{$(10^{-2} \mathrm{phot \cdot cm^{-2} \cdot}$} &
    \colhead{$(\mathrm{phot \cdot cm^{-2} \cdot}$} & 
    \colhead{$(10^{-2} \mathrm{phot \cdot cm^{-2} \cdot}$} &
    \colhead{$D_{norm}$} &
    \colhead{$D_{RS}$} \\
    \colhead{} &
    \colhead{($10^{-24}\,\mathrm{g \cdot cm^{-3}}$)} &
    \colhead{} &
    \colhead{($10^{22}\,\mathrm{cm^{-2}}$)} &
    \colhead{$\mathrm{s^{-1} \cdot keV^{-1}})$} &
    \colhead{$\mathrm{s^{-1} \cdot keV^{-1}})$} &
    \colhead{$\mathrm{s^{-1} \cdot keV^{-1}})$} &
    \colhead{(kpc)} &
    \colhead{(kpc)}
  }
  \startdata 
  DDTa & 2.0 & 0.01 & 0.94 & 2.97 & 2.79 & 8.3 & 3.58 & 2.57 \\
  DDTc & 2.0 & 0.03 & 0.55 & 2.89 & 1.78 & 8.1 & 4.48 & 2.59 \\
  DDTe & 2.0 & 0.1 & 0.36 & 2.81 & 1.36 & 7.9 & 5.12 & 2.54 \\
  5p0z22.25 & 5.0 & 0.05 & 0.57 & 2.89 & 0.74 & 8.1 & 6.95 & 1.95 \\
  PDDa & 0.2 & 0.005 & 0.76 & 2.81 & 68.3 & 7.9 & 0.72 & 4.38 \\
  W7 & 0.5 & 0.1 & 1.07 & 2.15 & 6.37 & 6.0 & 2.36 & 3.54 \\
  SCH & 0.5 & 0.01 & 1.03 & 2.60 & 10.82 & 7.3 & 1.82 & 3.44 \\
  b30\_3d\_768 & 0.2 & 0.01 & 2.33 & 0.35 & 59.8 & 1.0 & 0.77 & 3.41
  \enddata
  
  \tablenotetext{a}{The spectral approximations presented in Fig. \ref{fig-7} and in this Table are not statistically
    valid fits (see text), which makes it hard to define confidence ranges for the adjusted parameters $N_{H}$,
    $norm_{AM}$, and $norm_{ej}$. We do not provide any estimates on the errors for these parameters, but note that they
    may be large.}
 
\end{deluxetable}

\end{document}